%% file: main.tex
\documentclass{arxiv}
\biboptions{sort&compress}

\usepackage{amssymb}
\usepackage{amsmath}
\usepackage{booktabs}
\usepackage{caption}
\usepackage{subcaption}
\usepackage{xcolor}
\usepackage{float}
\usepackage{microtype}
\usepackage{hyperref}
\usepackage{xurl}%
\usepackage{tabularx}
\usepackage{array}
\usepackage{lineno}%

\hypersetup{
  colorlinks=true,
  linkcolor=black,
  citecolor=black,
  urlcolor=blue,
}

\begin{document}

\begin{frontmatter}

\title{Citywide building structural system mapping from sparse field data using satellite data and machine learning}

\author[1]{Javier Sempere-Hernández}
\author[2]{Miguel Ureña-Pliego\corref{cor1}}
\author[3]{Beatriz González-Rodrigo}
\author[4]{Beatriz Moya}
\author[4]{Sebastian Rodriguez}
\author[4,5]{Francisco Chinesta}
\author[2]{Miguel Marchamalo-Sacristán}

\cortext[cor1]{Corresponding author: Miguel Ure\~na-Pliego, miguel.urena@upm.es.}

\affiliation[1]{
  organization={Facultad de Ciencias,
    Universidad Nacional de Educación a Distancia},
  city={Madrid},
  country={Spain}
}
\affiliation[2]{
  organization={Escuela Técnica Superior de Ingenieros de Caminos,
    Canales y Puertos, Universidad Politécnica de Madrid},
  city={Madrid},
  country={Spain}
}
\affiliation[3]{
  organization={Escuela Técnica Superior de Ingenieros de Montes,
    Forestales y Medio Natural, Universidad Politécnica de Madrid},
  city={Madrid},
  country={Spain}
}

\affiliation[4]{
  organization={PIMM Laboratory,
Arts et Métiers Institute of Technology},
  city={Paris},
  country={France}
}
\affiliation[5]{
  organization={CNRS@CREATE LTD,
1 Create Way, \#08-01 CREATE Tower},
  country={Singapore}
}

\begin{abstract}
Seismic exposure inventories are essential bottlenecks for accurate seismic risk assessment and are often incomplete in many regions of the world where building information is scarce or not digitized. This study benchmarks and selects the best machine learning classification model to infer the structural system at the individual building scale from accessible data in three Latin American cities: Guatemala City, San Jos\'e (Costa Rica), and Santo Domingo (Dominican Republic). For this purpose, worldwide available geospatial data is employed, deriving cost-effective attributes from remote sensing and footprint geometry. Thus, transferability is assessed within and between cities and countries. Within-city accuracy varies widely between cities, mainly with how balanced and how distinct the structural classes of each city are. Transfer to other cities and districts without local training labels outperforms majority-class baselines but is mostly less effective than within-city prediction, and it cannot predict structural systems absent from the training data, such as adobe in Guatemala City. A single model trained on the combined data of the three cities performs close to the city-specific models, and labeling more buildings in a city improves its model only slightly once a few hundred labels are available. These findings show the potential of machine learning based on globally available geospatial attributes to expand seismic exposure inventories, while highlighting the need for local validation.
\end{abstract}

\begin{highlights}
\item Building structural systems inferred from footprints and open geospatial data
\item Models transferred to new cities perform moderately worse than local ones
\item A single model for three cities performs close to the city-specific models
\item With only a few hundred labeled buildings, a model reaches most of its accuracy
\item Model disagreement flags uncertain buildings to check or survey first
\end{highlights}

\begin{keyword}
seismic exposure model \sep building footprints \sep lateral load-resisting system \sep machine learning \sep geospatial data \sep spatial transferability
\end{keyword}

\end{frontmatter}
\linenumbers%

\section{Introduction}
\label{sec:introduction}
\input{sections/intro}

\section{Related Work}
\label{sec:related-work}
\input{sections/related-work}

\section{Dataset preparation}
\label{sec:data}
\input{sections/data}

\section{Methods}
\label{sec:methods}
\input{sections/methods}

\section{Experiments and Results}
\label{sec:results}
\input{sections/results}

\section{Conclusions}
\label{sec:conclusion}
\input{sections/conclusion}

\section{Future Work}
\label{sec:future-work}
\input{sections/future-work}

\section*{CRediT authorship contribution statement}
\textbf{Javier Sempere-Hern\'andez:} Conceptualization, Data curation, Formal analysis, Investigation, Methodology, Software, Visualization, Writing -- original draft, Writing -- review \& editing.
\textbf{Miguel Ure\~na-Pliego:} Conceptualization, Data curation, Formal analysis, Investigation, Methodology, Software, Validation, Visualization, Writing -- original draft.
\textbf{Beatriz Gonz\'alez-Rodrigo:} Conceptualization, Formal analysis, Funding acquisition, Investigation, Methodology, Project administration, Resources, Supervision, Writing -- original draft.
\textbf{Beatriz Moya:} Conceptualization, Investigation, Methodology, Software, Supervision, Validation, Writing -- review \& editing.
\textbf{Sebastian Rodriguez:} Investigation, Methodology, Software, Supervision, Validation, Writing -- review \& editing.
\textbf{Miguel Marchamalo-Sacrist\'an:} Conceptualization, Funding acquisition, Project administration, Resources, Supervision, Writing -- review \& editing.
\textbf{Francisco Chinesta:} Conceptualization, Project administration, Resources, Supervision.

\section*{Declaration of competing interest}
The authors declare that they have no known competing financial interests or personal relationships that could have appeared to influence the work reported in this paper.

\section*{Declaration of generative AI and AI-assisted technologies in the writing process}
During the preparation of this work the authors used Claude Sonnet 5.5 (Anthropic) for coding assistance, including the generation of better documentation, and for AI-assisted writing. The authors are not native English speakers, and this assistance enabled them to express themselves better. No AI tool was used for the first draft of the paper, for the literature review, or for reading the literature. After using this tool, the authors reviewed and edited the content as needed and take full responsibility for the content of the publication.

\section*{Funding}
This work was supported by the Spanish State Research Agency (AEI) of the Ministry of Science, Innovation and Universities under project RISK-CARIBERIA [PID2024-155792OB-I00], within the 2025 “Knowledge Generation” Call of the Spanish State Plan for Scientific, Technical and Innovation Research.

\section*{Data and code availability}
The data and code of this study are openly available under the European Union Public Licence v1.2 (EUPL-1.2), so every result can be reproduced from scratch.
{\urlstyle{same}
\begin{itemize}
  \item \textbf{Paper repository.} The six labeled surveys, the experiment configurations, the trained models and the analysis code.\\
  {\footnotesize \url{https://github.com/GeomaticsCaminosUPM/ml_structural_system_paper}}
  \item \textbf{Interactive map.} The surveyed buildings, the model predictions and their errors.\\
  {\footnotesize \url{https://geomaticscaminosupm.github.io/ml_structural_system_paper/map/}}
  \item \textbf{Classification method.} The \texttt{ml\_structural\_system} Python package and its documentation.\\
  {\footnotesize \url{https://github.com/GeomaticsCaminosUPM/ml_structural_system}}\\
  {\footnotesize \url{https://geomaticscaminosupm.github.io/ml_structural_system/}}
  \item \textbf{Footprint features.} The \texttt{footprint\_attributes} Python package, version 1.1, and its documentation.\\
  {\footnotesize \url{https://github.com/GeomaticsCaminosUPM/footprint_attributes}}\\
  {\footnotesize \url{https://geomaticscaminosupm.github.io/footprint_attributes/}}
\end{itemize}}
In the paper repository, a single script (\texttt{reproduce\_paper.py}) splits the data, trains and tests every model, and regenerates every number, table and figure reported here, with all package versions pinned to those used for this study. XGBoost and CatBoost were trained on a graphics processing unit, so retraining them on different hardware can change their scores slightly.

\section*{Acknowledgements}
The authors express their gratitude to Professors Diego Hidalgo-Leiva, Claudia Germoso and Omar Flores, and to the students who participated in field data collection.

\bibliography{references}

\end{document}

%% file: sections/intro.tex
Urban-scale seismic risk assessment requires a detailed characterization of the building stock through an exposure inventory, a database with an individual record for each building and the attributes that govern its structural vulnerability \citep{silva2022buildingclassification}. This information has traditionally been collected through field inspections, documentary reviews, and manual surveys, which demand considerable time, staff, and funding \citep{wieland2012exposure}, and whose labels depend on the judgment of each surveyor. These constraints prevent regular updates of exposure inventories and, in many contexts \citep{pittore2016exposure}, force the use of highly aggregated data or incomplete characterizations of the building stock \citep{torres2023remotesensing,yepesestrada2023global}. The quality of the exposure model directly affects the reliability of the risk assessment. Building-level detail allows a more accurate assignment of capacity curves and fragility or vulnerability functions, reduces the uncertainty of damage estimates, and supports loss estimation \citep{dabbeek2021exposure}, urban planning, and the prioritization of retrofitting, particularly where cadastral data are incomplete, outdated, or lack structural information.\\
Geospatial technologies, remote sensing, and artificial intelligence (AI) make it possible to build Geospatial Exposure Databases (GEDs) at the building level without inspecting every building. Orthophotos, light detection and ranging (LiDAR) point clouds, digital surface models, and drone imagery, combined with computer vision and machine learning, can automate footprint extraction, height estimation, and the identification of geometric and typological characteristics. In Port-au-Prince, \citet{torres2023remotesensing} built an exposure database from orthophotos and LiDAR data that reduced cost by up to 75\% and survey time by 85\% compared with field surveys, and the earthquake damage estimated with it closely matched that of a field-based reference database (mean absolute percentage error of 3.78\% for buildings with regular footprints).\\
Subsequent research has progressed toward scalable geospatial methodologies for data-scarce environments. \citet{rodriguezsaiz2025geospatial} combined multiple geospatial data sources to characterize seismic exposure at the individual-building level in San José, Costa Rica. \citet{urenapliego2026footprint} subsequently developed an AI-assisted procedure for footprint digitization and the automated extraction of geometry-derived seismic attributes and behavior modifiers, including building direction, position within the urban block (block position), and plan irregularity. These studies demonstrate that building footprints and other geospatial products can provide structured, reproducible, and spatially continuous information. However, they also show that not all attributes relevant to seismic response can be directly observed from aerial imagery.

The Global Earthquake Model (GEM) Building Taxonomy v2.0 defines 13 attributes related to the seismic performance of buildings \citep{brzev2013gemtaxonomy}. The typologies used to assign capacity and fragility curves are commonly defined by three of them, the material of the lateral load-resisting system (MLLRS), the lateral load-resisting system (LLRS), and height. These are among the factors with the greatest influence on seismic performance. In a worldwide expert survey, the structural system, its ductility, and height received the highest weights for comparing the seismic vulnerability of building classes \citep{jimenezmartinez2024fragility}, and their relative importance depends on the type of structure. The three attributes differ in how easily they can be obtained. Height or number of stories can be estimated from point clouds, digital surface models, or photogrammetric products. The ductility or seismic design level of the LLRS can be approximated from the construction period and the building codes in force at that time, although the construction year alone does not identify the load-resisting system unambiguously. The MLLRS, such as masonry, reinforced concrete, steel, or timber, is generally not expressed in the footprint geometry or in the roof visible in an orthophoto, so its determination at the urban scale typically requires facade observations, cadastral or documentary information, field inspections, or indirect inference.\\
Street-level imagery has supported the development of machine-learning models capable of recognizing typological attributes visible on building facades \citep{cardellicchio2022vulma}. However, services such as Google Street View, Mapillary, and comparable platforms do not provide spatially uniform coverage. This limitation is particularly relevant in parts of Central America and the Caribbean, where missing or outdated imagery, visual obstructions, and access restrictions prevent the systematic use of these platforms for comprehensive building inventories. Methods capable of inferring structural attributes that cannot be observed directly are therefore still needed, particularly approaches that exploit geospatial data sources with broader spatial coverage together with auxiliary variables associated with individual buildings.\\

This study develops a machine learning pipeline to train and test classification algorithms that infer the MLLRS of individual buildings from geospatial information and observable or derived attributes. For brevity, the MLLRS, the material of the lateral load-resisting system, is referred to as the structural system in the rest of the paper. It is applied to three pilot areas in San José (Costa Rica), Guatemala City (Guatemala), and Santo Domingo (Dominican Republic), urban contexts of Central America and the Caribbean where street-level imagery is limited or spatially inconsistent. The predictors are mainly features derived from building footprints and their spatial relationships with neighboring buildings, complemented by inexpensive and commonly available attributes such as height and approximate construction period. None of them observes the structural system directly, and masonry accounts for more than 80\% of the labeled buildings in most data subsets, so the task is an imbalanced classification problem based on indirect indicators. In particular, five classification algorithms are compared, from a linear baseline to gradient-boosted tree ensembles (logistic regression, random forest, XGBoost, CatBoost and histogram-based gradient boosting). They are evaluated with spatially grouped partitions within each city, across cities and districts without target labels, and as a single model trained on the combined data of the three cities, together with the number of local labels required and the most informative features. The main contribution is an empirical evaluation of an automated and scalable structural-system inference pipeline on real-world, unevenly labeled, multi-city data, quantifying how far the resulting models generalize beyond the buildings, districts, and cities represented in their training data.\\
The paper is structured as follows. Section~\ref{sec:related-work} reviews previous work on inferring structural systems from remote sensing and geospatial data. Section~\ref{sec:data} describes the surveys and the variables derived from footprints and remote-sensing products. Section~\ref{sec:methods} presents the classification pipeline, the handling of class imbalance, the models and the evaluation metrics. Section~\ref{sec:results} reports the results within cities, the training-data requirements, the transfer between cities and districts, the three-city model and the feature importance. Section~\ref{sec:conclusion} presents the conclusions and the limitations of the study, and Section~\ref{sec:future-work} outlines future work.

%% file: sections/related-work.tex
Inferring structural systems from data sources less costly than in-person
surveys is an established approach in earthquake engineering, using satellite, aerial and ground-based remote sensing \citep{pittore2013toward,geiss2015estimation,wieland2012exposure}, rule learning and support vector machines on census data \citep{riedel2015seismic}, street-level imagery \citep{aravenapelizari2021automated,cardellicchio2022vulma}, and multi-source geospatial databases \citep{torres2023remotesensing,rodriguezsaiz2025geospatial}. Regarding the transferability questions examined in this study, \citet{liuzzi2019transferable} classified 17462 buildings of 18 villages in the Val d'Agri area, Italy, as masonry or reinforced concrete with a Random Forest classifier, using features derived from building footprints, height and a 50\,cm aerial orthophoto. Almost all of the ten most important features were geometric, such as area, perimeter, height and shape, and the authors kept them because they are less sensitive to differences between images. The model was trained in one village (Paterno, 1844 buildings) and transferred to the other 17 villages (15618 buildings). With the class weighting used for transfer, Cohen's $\kappa$ went from 0.36 within Paterno to 0.31 after transfer, a fair agreement between predicted and observed classes, with a True Skill Statistic (TSS) of 0.35 for the ability to discriminate between classes \citep{allouche2006assessing}, both defined in Section~\ref{sec:evaluation}. In individual villages, $\kappa$ ranged from 0.36 to 0.62. The authors considered these results encouraging and noted that further improvement would require additional knowledge of the target domain. With respect to the feature-engineering
approach adopted here, \citet{geiss2015estimation} classified seismic building
structural types in Padang, Indonesia, using building footprint, height information, and
multi-temporal remote-sensing features with Support Vector Machine (SVM) and Random Forest classifiers. The study faced similar scarce and imbalanced labels, comprising 573 buildings and likewise employed
Synthetic Minority Over-sampling Technique (SMOTE) resampling \citep{chawla2002smote}. However, the authors noted that cross-validation accuracy obtained after SMOTE-based augmentation may be overly optimistic. For this reason, all performance scores reported in the present study are calculated using held-out test data.

Building and urban structure types have also been classified from footprint geometry outside seismic vulnerability assessment, mainly in urban studies and cartography, using footprints from cadastral and topographic databases \citep{steiniger2008classification,hecht2015buildingtypes}, 3D city models \citep{henn2012automatic} and OpenStreetMap \citep{fan2014estimation,atwal2022predicting}. \citet{hecht2015buildingtypes} and \citet{atwal2022predicting} distinguish residential and non-residential building types and reach high accuracies in the areas where they were trained. Transfer works between areas with a similar building culture, as among three United States study areas \citep{atwal2022predicting}, but is otherwise limited by regional differences in urban structure \citep{steiniger2008classification,hecht2015buildingtypes}. \citet{aravenapelizari2021automated} inferred comparable structural types from street-level imagery using deep
learning methods instead of footprint geometry. That study identified two limitations that are also relevant here: labels obtained through visual interpretation cannot be considered verified ground
truth, and transferring a model to a new geographical region may require adapting the building typologies to the local construction context.

Instead of working with images, the present study uses features derived from building footprints, namely plan-irregularity indices adapted from five national and international seismic codes (compactness, eccentricity, setback, slenderness and parallelity), the position of the building within its block, and its roof material, computed with the open-source pipeline of \citet{urenapliego2026footprint}. Seismic codes regulate plan irregularity because it affects seismic behavior, the position within the block determines how neighbors restrain or push a building, and all of them can be computed from footprints available worldwide. The present study extends the
city-scale Geospatial Exposure Database methodology by
\citet{rodriguezsaiz2025geospatial}, developed for San Jos\'e, to
cross-city and cross-district transfer scenarios. That study classified
7226 San Jos\'e buildings into eight GEM-aligned typologies using a Random Forest classifier and the multi-hazard building classification system proposed by
\citet{silva2022buildingclassification}. Its classification error
decreased from 46\% to 13\% when visually estimated height was included. The two studies address different questions. \citet{rodriguezsaiz2025geospatial} predict eight typologies within a single city from a large labeled sample, whereas the present study reduces these to four classes shared across the three cities and, because its focus is transfer to areas without local labels, evaluates every model on a held-out, spatially grouped test partition.

Spatially aware data partitioning is applied to ensure that the test set reflects the spatial dependence present in the data. \citet{salazar2022fairsplit} employed kriging variance as a proxy for spatial prediction difficulty, whereas \citet{wang2023spatialplus} combined hierarchical-clustering blocks with an assessment of similarity in feature space. Both approaches produce evaluation partitions that more closely represent real deployment conditions than conventional random splitting. The present study adopts a simpler strategy in which physically contiguous blocks are kept intact and assigned entirely to the training, validation, or test set. This approach is considered sufficient because the principal risk of data leakage is direct (neighboring or touching buildings may share footprint-derived features) rather than being associated with a broader gradient in spatial prediction difficulty.

%% file: sections/data.tex
The study combines two types of data. Field surveys provide the structural system of each building, the target of the classification, together with its location and, in five of the six surveys, its number of stories. Building footprints, complemented by remote-sensing products, provide every predictor variable. This section describes the surveyed areas, the variables derived from footprints and remote sensing, and the distribution of structural classes.

\subsection{Areas under study for survey collection}
\label{sec:data-cities}

\begin{figure}[H]
\centering
\includegraphics[width=\textwidth]{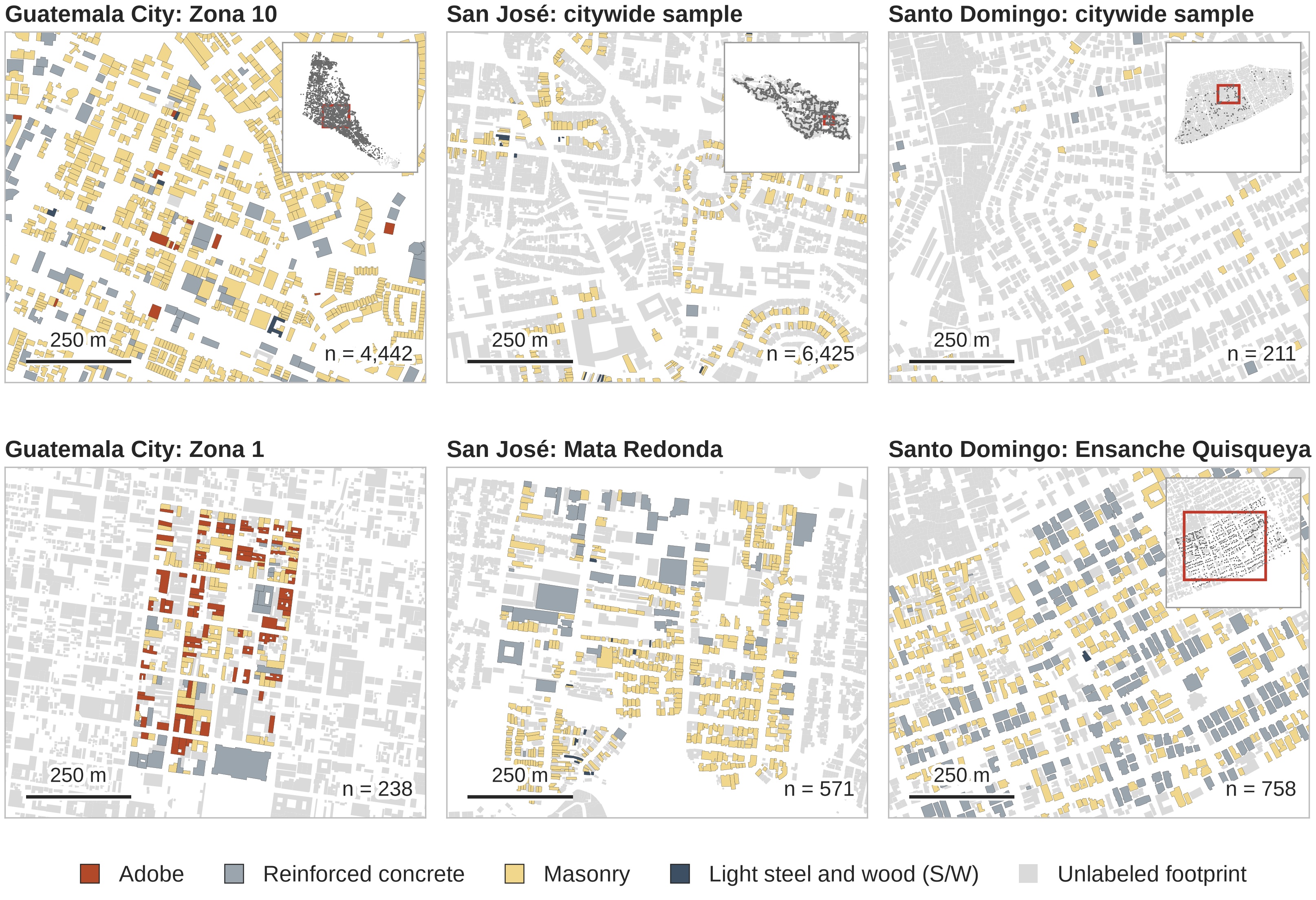}
\caption{Labeled buildings of the six surveys over unlabeled footprints (gray). Top: larger-area survey of each city; bottom: single-district survey; $n$: number of labeled buildings. Background map: Microsoft Building Footprints database and Overture Maps Foundation.}
\label{fig:data-collage}
\end{figure}

The survey dataset comprises six independently labeled surveys distributed across three cities. In Guatemala City, the dataset includes a localized survey within Zona 1, the central district of the city,  \citep{gonzalez2024sustainable} and a larger survey covering Zona 10 \citep{davila2023guatemala}. In San José, the dataset includes a localized survey within the Mata Redonda district \citep{gonzalez2024sustainable} and a substantially larger survey that encompasses multiple districts (Catedral, Carmen, Hatillo, Hospital, Mata Redonda, Merced, Pavas, San Francisco de Dos Ríos, San Sebastián, Uruca, and Zapote) \citep{esquivelsalas2021sanjose} and whose extent includes the Mata Redonda survey area, hereafter referred to as the San José citywide sample. In Santo Domingo, the dataset consists of a localized survey within Ensanche Quisqueya \citep{gonzalez2024sustainable} and a survey covering a larger area, concentrated in Julieta Morales and neighboring sectors, hereafter referred to as the Santo Domingo citywide sample. Unlike the other two cities, this larger-area survey contains fewer labeled buildings (211) than the localized one (758). The spatial distribution of all six surveys is presented in Figure~\ref{fig:data-collage}. Within each city, the localized survey and its corresponding large-scale counterpart were digitized and labeled independently, providing distinct datasets for model development and evaluation. Because the surveys overlap geographically in San José and Santo Domingo, a small number of physical buildings appear in both surveys of the same city (independently digitized footprints with an intersection over union of at least 0.5): 31 buildings in San José (0.5\% of the labeled citywide sample) and 11 in Santo Domingo (5.2\% of the citywide sample), of which 28 and 10, respectively, carry the same structural class in both surveys. These buildings were retained in both surveys.

\subsection{Feature extraction from footprint information}
\label{sec:data-features}

\begin{figure}[H]
\centering
\includegraphics[width=\textwidth]{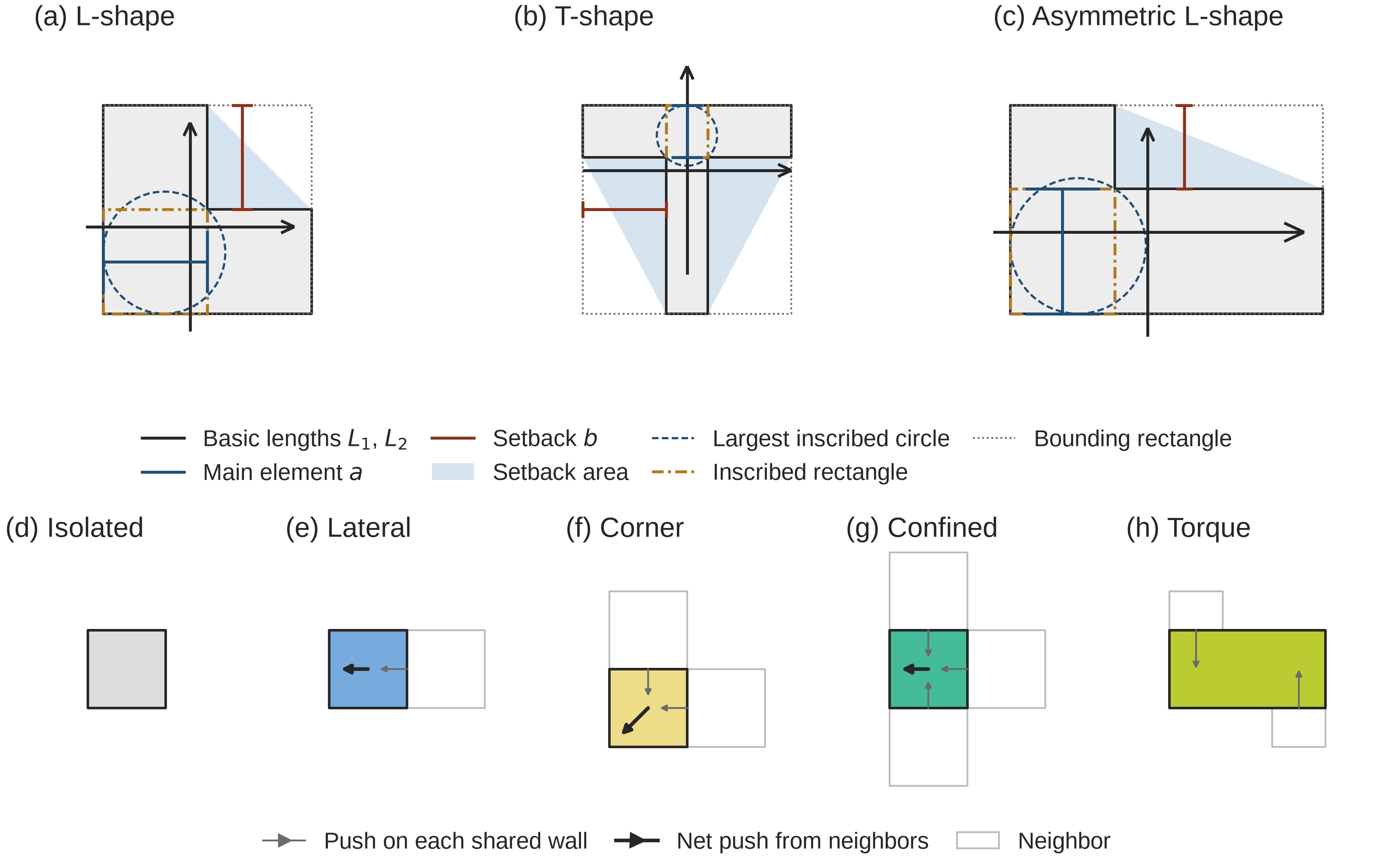}
\caption{Footprint descriptors on simple shapes. (a--c) Building sizes behind the plan-irregularity indices. (d--h) Block positions from the contact-force analogy.}
\label{fig:footprint-descriptors}
\end{figure}

The predictor variables were derived primarily from the footprint of each building and its spatial
relationship with the adjacent structures using the open-source
footprint-processing workflow developed by \citet{urenapliego2026footprint}. The workflow takes each footprint and its neighbors as input. Plan-shape indices are computed from the polygon alone, by comparing it with reference shapes such as its convex hull, its minimum bounding rectangle and the circle of equal area, following the definitions of each seismic code (Figure~\ref{fig:footprint-descriptors}a--c). Neighbor descriptors use a contact-force analogy. A virtual pressure acts on every wall shared with an adjacent building, giving a force proportional to the wall length and the building height. The magnitude of the net force, the share of it opposed by walls on the opposite side and the spread of the wall directions classify each building as isolated, lateral, corner, confined or torque (Figure~\ref{fig:footprint-descriptors}d--h). Walls closer than 0.1\,m are treated as shared, to absorb small digitization gaps.

Three additional attributes, selected for their global availability and low acquisition cost, were obtained from remote
sensing products rather than directly from the footprint geometry. Building height was taken from the 
GlobalBuildingAtlas \citep{zhu2025globalbuildingatlas}, replaced by drone measurements or by the surveyed number of stories (3\,m per story) where these disagreed with it, while the approximate construction era was inferred from the World Settlement Footprint Evolution \citep{marconcini2021wsfevolution} and Global Human Settlement built-up surface (GHS-BUILT-S) products
 \citep{pesaresi2024ghsl}, updated with the most recent change detected in Landsat and Sentinel-2 time series. Roof material was determined at the building level through the fusion of Sentinel-2 imagery and high-resolution orthophotographs. Notably, none of these datasets provides direct information on the structural system.\\ 
A total of \textbf{27 variables for each building} were considered. The categorical descriptors of roof material and block position were one-hot encoded with one reference category dropped, so the number of input variables depends on the categories observed in each city's training data: 31 in Guatemala City and Santo Domingo, and 32 in San Jos\'e. These variables fall into three groups:

\begin{itemize}
  \item \textbf{Remote-sensing attributes}: height, construction era, roof material, and a seismic code-quality level derived from the era and each country's seismic-code years (Costa Rica: 1974, 1986, 2003; Guatemala: 1996; Dominican Republic: 1979, 2011).
  \item \textbf{Plan-shape and irregularity descriptors}: compactness, convex-hull irregularity, inertia-to-circle ratio, radius ratio, and slenderness, eccentricity, setback, hole, and parallelity indices adapted from the irregularity provisions of Eurocode 8, the American Society of Civil Engineers standard ASCE 7, the Italian National Group for Earthquake Defence level-II form (GNDT-II), the Seismic Code of Costa Rica 2010 (CSCR-2010), and the Mexico City Complementary Technical Standards 2023 (NTC-23), computed following \cite{urenapliego2026footprint}.
  \item \textbf{Neighbor-interaction descriptors}: contact forces with adjacent buildings (magnitude, confinement, and orientation) and the resulting block position (isolated, lateral, corner, confined, or torque), computed following \citet{urenapliego2026footprint}.
\end{itemize}

\begin{figure}[H]
\centering
\includegraphics[width=\textwidth]{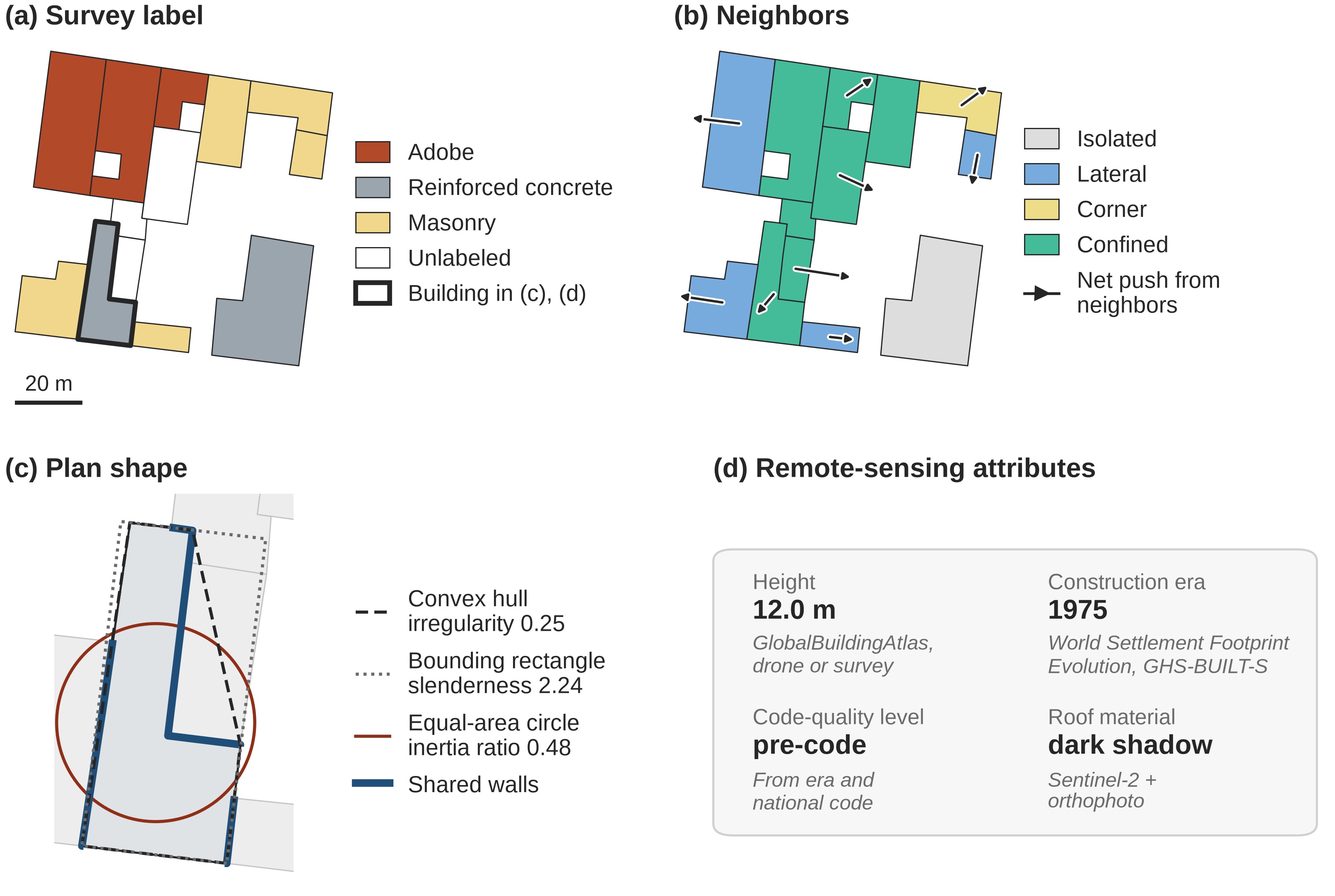}
\caption{Model input for one block of Zona 1, Guatemala City. (a) Survey label, the prediction target. (b) Block position and net contact forces. (c) Plan-shape and contact descriptors and (d) remote-sensing attributes of the outlined building.}
\label{fig:model-input}
\end{figure}

Figure~\ref{fig:model-input} illustrates these attributes for one block of Zona 1, Guatemala City. None of them observes the structural system directly.

The surveyed building stock exhibited limited structural-typology diversity. Consequently, the original labels were aggregated into four broad structural classes, adobe (ADO), reinforced concrete (CR), masonry (M), and light steel and wood (S/W). Steel frames were grouped with reinforced concrete, and light steel systems were merged with light wood systems into S/W, because both are structurally similar lightweight framed systems and the merged class is larger. Although this classification is coarser than the eight GEM-compatible expert-defined building typologies identified for San José by \citet{rodriguezsaiz2025geospatial}, it provides a suitable level of resolution for the inter-city and inter-district transferability analyses presented below.
\label{sec:data-imbalance}

\begin{table}[H]
\centering
\small
\caption{Labeled buildings per city (row percentages). ADO: Adobe; CR:
Reinforced Concrete; M: Masonry; S/W: Light steel and wood.}
\label{tab:class-imbalance}
\begin{tabular}{lrrrrr}
\toprule
City & $N$ & ADO & CR & M & S/W \\
\midrule
Guatemala City & 4680 & 174 (4\%) & 492 (11\%) & 3991 (85\%) & 23 (0\%) \\
San Jos\'e & 6996 & 2 (0\%) & 294 (4\%) & 6440 (92\%) & 260 (4\%) \\
Santo Domingo & 969 & 0 (0\%) & 307 (32\%) & 660 (68\%) & 2 (0\%) \\
\midrule
All three cities & 12645 & 176 (1\%) & 1093 (9\%) & 11091 (88\%) & 285 (2\%) \\
\bottomrule
\end{tabular}
\end{table}

Masonry is the predominant structural class in all three city datasets, accounting for 68\% of buildings in Santo Domingo, the smallest dataset and the
most balanced in terms of class distribution, and 92\% in San Jos\'e. Light steel and wood buildings (S/W) occur almost exclusively in San Jos\'e, whereas adobe buildings are largely confined to Guatemala City. Both constitute minority classes within their respective city datasets. This imbalance is a central difficulty of the task, because a classifier that predicts masonry for every building reaches high accuracy while missing the minority classes, such as adobe, that matter most for vulnerability.

%% file: sections/methods.tex
\subsection{Pipeline}
\label{sec:pipeline}

\begin{figure}[H]
\centering
\includegraphics[width=\textwidth]{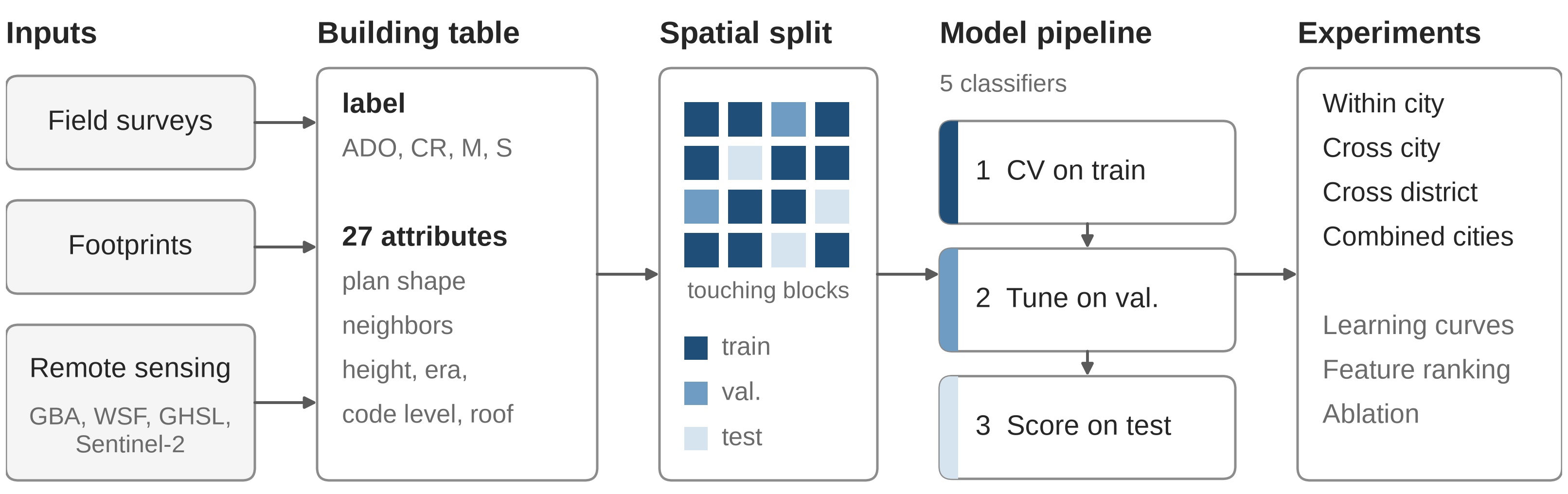}
\caption{Study workflow, with inputs from GlobalBuildingAtlas (GBA), World Settlement Footprint Evolution (WSF), and GHS-BUILT-S from the Global Human Settlement Layer (GHSL). Partitions are formed from blocks of touching buildings.}
\label{fig:workflow}
\end{figure}

This work evaluates five state-of-the-art classification algorithms to provide a broad overview of the efficiency of the features selected for the inference of the structural system of individual buildings. Figure~\ref{fig:workflow} summarizes the complete workflow. The labeled buildings of each experiment are divided into training, validation, and test partitions using spatially grouped splits. All models then follow the same three-stage workflow: configuration selection, decision-rule adjustment, and final evaluation. First, a grid search with stratified five-fold
cross-validation on the training partition evaluates hyperparameter configurations and, independently,
the imbalance-handling strategies described below. Configurations are ranked by their mean macro F1 score across folds. 
Second, the model with the selected configuration is fitted on the training partition, and its decision rule is adjusted on the validation partition. Class-specific probability divisors are optimized to maximize macro F1 on the validation partition, using a coordinate-wise grid search over each class's divisor, and predictions are obtained by selecting the class with the highest \emph{adjusted} probability. This adjustment addresses the tendency of the standard maximum-probability decision rule to favor the majority masonry class, which accounts for 68--92\% of buildings depending on the city. Decision-rule adjustment complements class weighting and resampling: these strategies influence model fitting, whereas the probability divisors modify the subsequent assignment of class labels.\\
Finally, the fitted model and adjusted decision rule are evaluated once on a held-out, spatially grouped test set. This test set is excluded from configuration selection, decision-rule adjustment, and model fitting. The validation partition is used only for decision-rule adjustment.

\subsection{Spatially grouped partitions}
\label{sec:space-aware}

A random building-level split can overestimate performance when spatial dependence links training and test samples \citep{roberts2017crossvalidation}, and several predictors here describe each building's physical neighborhood. The training, validation, and test partitions are therefore spatially grouped: buildings that physically touch are merged into contiguous blocks, and each block is assigned entirely to one partition.

\subsection{Handling class imbalance}
\label{sec:imbalance}

Resampling methods address the relative imbalance between classes but cannot compensate for rare classes whose examples were never collected \citep{he2009imbalanced}. This distinction is particularly important when a class is rare because it is geographically concentrated. For example, if adobe buildings occur almost exclusively in Guatemala City within the study data, no resampling strategy can compensate for its absence in areas where representative observations are unavailable.\\
The cross-validation grid also compares the following imbalance-handling strategies for each model and training
set: no resampling, inverse-frequency class weighting, and SMOTE
\citep{chawla2002smote}, optionally followed by Tomek-link
cleaning \citep{tomek1976cnn}. The latter two procedures are implemented using imbalanced-learn \citep{lemaitre2017imblearn}, and strategies are evaluated and selected based on mean macro F1. Random Forest already under-samples the majority
class for each tree, so only the no-resampling and class-weighting options
are evaluated for this model.

\subsection{Models}
\label{sec:models}

Five models are compared (Table~\ref{tab:ml_methods}), sharing the same cross-validation, imbalance handling, and decision-rule adjustment. L2-regularized logistic regression \citep{hastie2009esl} is the linear baseline. Random Forest is used in its balanced variant \citep{chen2004balancedrf}, which under-samples the majority class in each tree's bootstrap sample. XGBoost \citep{chen2016xgboost} and CatBoost \citep{prokhorenkova2018catboost} are gradient-boosted tree ensembles. CatBoost differs mainly in its ordered boosting, which reduces the prediction shift caused by target leakage. Histogram-based gradient boosting (Hist. GB), inspired by LightGBM \citep{ke2017lightgbm}, bins feature values to speed up training. All three boosting models use early stopping: XGBoost and CatBoost on the held-out fold during cross-validation, with the final model trained for the mean best iteration across folds, and histogram-based gradient boosting on an internal 10\% validation split.

\begin{table}[H]
\centering
\small
\caption{Overview of the machine learning models, their main
technical characteristics, and training strategies.}
\label{tab:ml_methods}
\footnotesize
\setlength{\tabcolsep}{3pt}
\renewcommand{\arraystretch}{1.3}
\begin{tabular}{@{}p{0.20\linewidth}p{0.46\linewidth}p{0.26\linewidth}@{}}
\toprule
Model & Technical characteristics & Training strategy \\
\midrule

Logistic Regression
\citep{hastie2009esl}
& Linear probabilistic classifier that estimates class
probabilities from weighted input features. L2
regularization limits coefficient magnitude and reduces
overfitting.
& Linear baseline with L2 regularization. \\

\midrule

Balanced Random Forest
\citep{chen2004balancedrf}
& Ensemble of independently trained decision trees.
Captures nonlinear relationships and feature interactions.
Majority classes are under-sampled in each tree's
bootstrap sample to address class imbalance.
& Balanced bootstrap sampling; predictions aggregated
across trees. \\

\midrule

XGBoost
\citep{chen2016xgboost}
& Gradient-boosted decision trees trained sequentially
to minimize a regularized loss function. Each tree
improves the predictions of the preceding ensemble.
& Early stopping on held-out cross-validation folds;
final iterations based on the mean best iteration. \\

\midrule

CatBoost
\citep{prokhorenkova2018catboost}
& Gradient-boosted decision trees with ordered boosting
to reduce prediction shift associated with target
leakage. Supports categorical feature processing.
& Early stopping on held-out cross-validation folds;
final iterations based on the mean best iteration. \\

\midrule

Hist.\ GB
\citep{ke2017lightgbm}
& Gradient-boosted decision trees using discretized
feature values (histograms) to accelerate split
selection and reduce computational cost.
& Early stopping using an internal 10\% validation
split. \\

\bottomrule
\end{tabular}
\end{table}

\subsection{Evaluation}
\label{sec:evaluation}

Four metrics are reported, macro F1 and accuracy \citep{sokolova2009systematic}, Cohen's $\kappa$ \citep{cohen1960coefficient}, and the macro-averaged True Skill Statistic (TSS) \citep{allouche2006assessing}. For each class $c$ of the $K$ classes in a test set of $N$ buildings, $TP_c$, $FP_c$, $FN_c$ and $TN_c$ count the true positives, false positives, false negatives and true negatives, giving precision $P_c = TP_c/(TP_c+FP_c)$, recall $R_c = TP_c/(TP_c+FN_c)$ and specificity $S_c = TN_c/(TN_c+FP_c)$. Macro F1 and accuracy follow Equation~\eqref{eq:f1-accuracy}, and Cohen's $\kappa$ and TSS follow Equation~\eqref{eq:kappa-tss}.
\begin{equation}
\label{eq:f1-accuracy}
\begin{gathered}
\text{macro F1} = \frac{1}{K}\sum_{c=1}^{K}\frac{2P_cR_c}{P_c+R_c}, \\
\text{accuracy} = p_o = \frac{1}{N}\sum_{c=1}^{K}TP_c,
\end{gathered}
\end{equation}
\begin{equation}
\label{eq:kappa-tss}
\begin{gathered}
\kappa = \frac{p_o-p_e}{1-p_e}, \\
\text{TSS} = \frac{1}{K}\sum_{c=1}^{K}\left(R_c+S_c-1\right),
\end{gathered}
\end{equation}
where $p_e = \sum_{c} n_c\,m_c/N^2$ is the agreement expected by chance from the numbers of buildings $n_c$ labeled and $m_c$ predicted as class $c$. Macro F1 is the headline metric because of the class imbalance. Macro averages use the classes present in each test set, and predictions of absent classes count as errors. Kappa and TSS are reported wherever a
test set has more than one class actually predicted. Feature
importance is a consensus
ranking across models, built from SHAP
values \citep{lundberg2017shap} for four of the five models and
permutation importance (macro F1 drop when a feature is shuffled) for
Histogram-based gradient boosting.

%% file: sections/results.tex
In this section, we report the results obtained for the five preselected classification algorithms for city-specific and global models. Every table below also reports a majority-class baseline, the score of always predicting the most common class, to assess whether the models provide meaningful predictive value beyond a trivial classification strategy, particularly in the presence of class imbalance.

\subsection{Base-city models}
\label{sec:exp-basecities}

In this section we showcase the results of the classifiers trained on their own datasets, without transfer of knowledge. Table~\ref{tab:basecities} reports all five models within each city, trained on 2954, 4604 and 619 buildings, with 778, 1036 and 121 for validation and 892, 1354 and 225 held out for testing in Guatemala City, San Jos\'e and Santo Domingo.

\begin{table}[H]
\centering
\small
\caption{Base-city models, best macro F1 per city in bold. 95\% CI: percentile bootstrap over 2000 resamples of the test set's spatial blocks.}
\label{tab:basecities}
\footnotesize
\begin{tabular}{llrrrrr}
\toprule
 & & \multicolumn{2}{c}{Macro F1} & & & \\
\cmidrule(lr){3-4}
City & Model & Value & 95\% CI & Acc. & $\kappa$ & TSS \\
\midrule
Guatemala City & CatBoost & \textbf{0.58} & [0.47, 0.69] & 0.86 & 0.41 & 0.38 \\
 & Hist.\ GB & 0.58 & [0.46, 0.68] & 0.88 & 0.42 & 0.34 \\
 & Logistic Regression & 0.55 & [0.47, 0.62] & 0.84 & 0.39 & 0.38 \\
 & Random Forest & 0.57 & [0.45, 0.67] & 0.86 & 0.38 & 0.34 \\
 & XGBoost & 0.56 & [0.45, 0.66] & 0.84 & 0.37 & 0.37 \\
 & \emph{Majority baseline} & \emph{0.31} & \emph{--} & \emph{0.87} & \emph{0.00} & \emph{0.00} \\
\midrule
San Jos\'e & CatBoost & \textbf{0.71} & [0.65, 0.77] & 0.94 & 0.58 & 0.61 \\
 & Hist.\ GB & 0.68 & [0.62, 0.74] & 0.94 & 0.53 & 0.54 \\
 & Logistic Regression & 0.63 & [0.57, 0.69] & 0.92 & 0.44 & 0.47 \\
 & Random Forest & 0.63 & [0.56, 0.70] & 0.94 & 0.50 & 0.47 \\
 & XGBoost & 0.69 & [0.63, 0.75] & 0.94 & 0.55 & 0.56 \\
 & \emph{Majority baseline} & \emph{0.32} & \emph{--} & \emph{0.93} & \emph{0.00} & \emph{0.00} \\
\midrule
Santo Domingo & CatBoost & 0.93 & [0.89, 0.96] & 0.94 & 0.86 & 0.86 \\
 & Hist.\ GB & 0.93 & [0.89, 0.96] & 0.94 & 0.86 & 0.89 \\
 & Logistic Regression & 0.93 & [0.88, 0.96] & 0.95 & 0.86 & 0.81 \\
 & Random Forest & \textbf{0.94} & [0.91, 0.97] & 0.96 & 0.89 & 0.89 \\
 & XGBoost & 0.94 & [0.90, 0.97] & 0.95 & 0.87 & 0.90 \\
 & \emph{Majority baseline} & \emph{0.42} & \emph{--} & \emph{0.72} & \emph{0.00} & \emph{0.00} \\
\bottomrule
\end{tabular}
\end{table}

\begin{figure}[H]
\centering
\includegraphics[width=\textwidth]{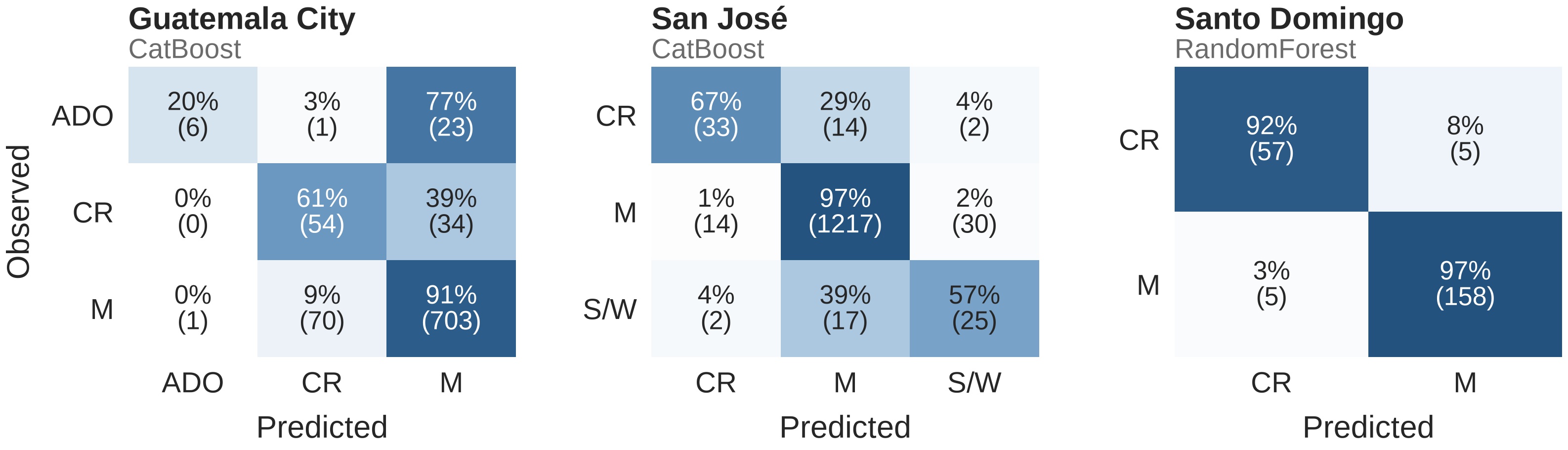}
\caption{Confusion matrix of each city's best model (named under the city), row-normalized to the percentage of each observed class, with building counts in parentheses. ADO: Adobe; CR: Reinforced Concrete; M: Masonry; S/W: Light steel and wood.}
\label{fig:exp1-confusion}
\end{figure}

Santo Domingo, the smallest and most balanced dataset (Table~\ref{tab:class-imbalance}), yields the highest performance (best $\kappa = 0.89$), whereas the best models reach $\kappa = 0.41$ in Guatemala City and $0.58$ in San Jos\'e, with frequent confusion between reinforced concrete and masonry (Figure~\ref{fig:exp1-confusion}). In San Jos\'e, S/W also has low recall (57\%). CatBoost leads in Guatemala City and San Jos\'e and random forest in Santo Domingo, but a paired block bootstrap of the macro-F1 difference with the second-ranked model includes zero in all three cities.

\label{sec:exp-whycity}
Errors shared by all five models affect 6.4\% of test buildings in Guatemala City, 1.7\% in San Jos\'e, and 0.9\% in Santo Domingo, mostly confusions between reinforced concrete and masonry in Guatemala City (38 of 57) and between S/W and masonry in San Jos\'e (14 of 23). Height separates reinforced concrete from masonry more clearly in Santo Domingo (Cohen's $d = 3.4$) than in Guatemala City ($d = 1.1$), consistent with its fewer shared errors, although part of this separation comes from drone-measured heights ($d = 1.5$ with GlobalBuildingAtlas heights alone). In San Jos\'e, S/W and masonry differ in height only through surveyed story counts ($d = 0.9$, versus $0.1$ with GlobalBuildingAtlas alone).

\subsection{Training-data requirements}
\label{sec:exp-learningcurve}

\begin{figure}[H]
\centering
\includegraphics[width=\textwidth]{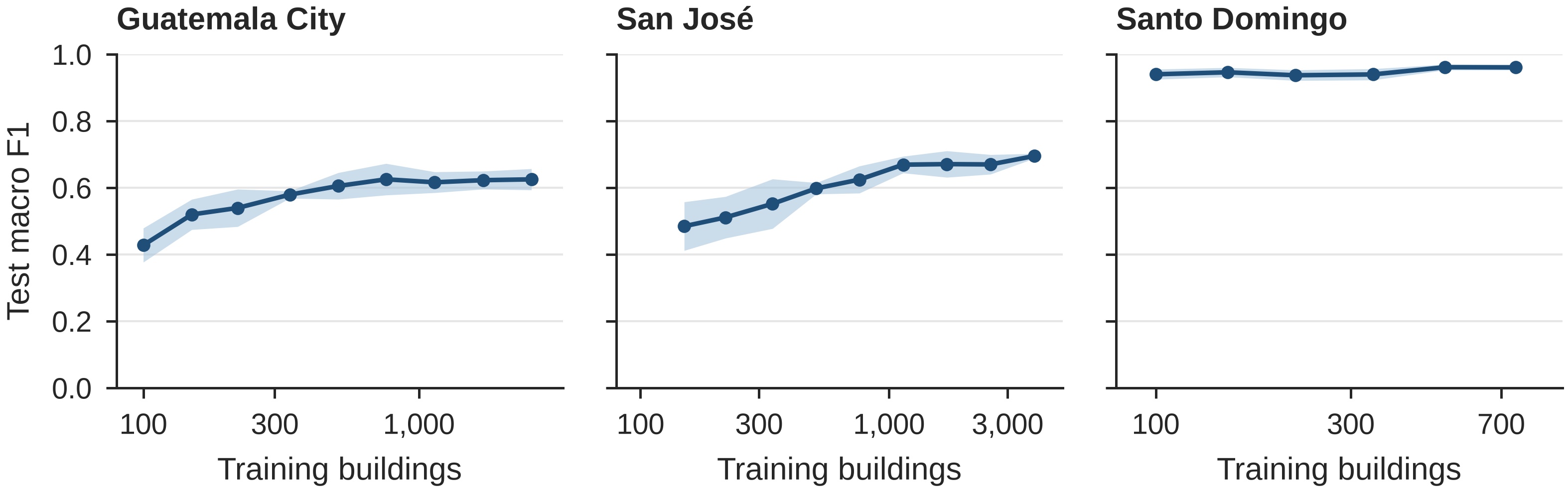}
\caption{Learning curves for XGBoost in Guatemala City, San José, and Santo Domingo, showing test macro F1 as a function of training-set size. Results are summarized as the mean $\pm$ one standard deviation across up to five repeated spatially grouped splits.}
\label{fig:learning-curve}
\end{figure}

Figure~\ref{fig:learning-curve} shows distinct patterns across cities. In Santo Domingo, macro F1 is already high with 100 training buildings (0.94) and increases by only +0.02 over an additional approximately eightfold increase in training-set size. This near-plateau suggests limited gains from additional labeled data within the evaluated range. In contrast, performance improves substantially in Guatemala City and San Jos\'e. In
Guatemala City, macro F1 increases from 0.43 with 100 training buildings to 0.62 with 2560 buildings, a gain of approximately +0.20. In San
Jos\'e, it increases from 0.48 with 150 buildings, the smallest size with at least three valid repeats, to 0.69 with 3840 buildings, a gain of approximately +0.21. In both cities the gains become small above about 760 (Guatemala City) and 1140 (San Jos\'e) training buildings, about +0.01 to +0.02 macro F1 per doubling of the training set, so further local labeling may still bring modest improvements in these two cities. About 100 training buildings in Santo Domingo and 340 to 510 in Guatemala City come within 0.05 macro F1 of the largest training sets. San Jos\'e, where S/W is rare, needs 760 to 1140.

\subsection{Cross-city and district transfer}
\label{sec:exp-cityloo}
\label{sec:exp-districttransfer}

Building on the within-city results (Table~\ref{tab:basecities}), we evaluate model transferability across cities and districts without using target-area labels for training. 
In the cross-city scenario, models are trained on two cities and evaluated on the third using a leave-one-city-out approach. In the within-city scenario, models are
trained on the single-district survey and evaluated on the survey covering a larger area of the same city. Together, these experiments assess the applicability of trained models to areas where local labeled data are unavailable for model development. Macro F1 is averaged over the classes present in each target test set, and the transfer and within-city models are compared on the classes shared by both test sets, reinforced concrete and masonry in every case. 

\begin{table}[H]
\centering
\small
\caption{Best model (highest macro F1) per transfer setting, with majority-class baselines. $\Delta$F1 is the difference from the best within-city model (Table~\ref{tab:basecities}) on the classes shared by both test sets. 95\% CI: percentile bootstrap over 2000 resamples of spatial blocks.}
\label{tab:transfer}
\footnotesize
\begin{tabular}{llrrrrr}
\toprule
 & & \multicolumn{2}{c}{Macro F1} & & & \\
\cmidrule(lr){3-4}
Transfer & Best model & Value & 95\% CI & $\Delta$F1 & $\kappa$ & TSS \\
\midrule
\multicolumn{7}{l}{\textit{Leave-one-city-out (trained on the other two cities)}} \\
Guatemala City & Random Forest & 0.73 & [0.70, 0.76] & $+0.02$ & 0.46 & 0.40 \\
 & \emph{Majority baseline} & \emph{0.47} & \emph{--} & \emph{--} & \emph{0.00} & \emph{0.00} \\
San Jos\'e & Logistic Regression & 0.48 & [0.46, 0.49] & $-0.11$ & 0.26 & 0.19 \\
 & \emph{Majority baseline} & \emph{0.33} & \emph{--} & \emph{--} & \emph{0.00} & \emph{0.00} \\
Santo Domingo & XGBoost & 0.57 & [0.56, 0.59] & $-0.08$ & 0.72 & 0.49 \\
 & \emph{Majority baseline} & \emph{0.27} & \emph{--} & \emph{--} & \emph{0.00} & \emph{0.00} \\
\midrule
\multicolumn{7}{l}{\textit{Single-district survey $\to$ larger-area survey, same city}} \\
Guatemala City & XGBoost & 0.69 & [0.66, 0.71] & $-0.02$ & 0.38 & 0.32 \\
 & \emph{Majority baseline} & \emph{0.47} & \emph{--} & \emph{--} & \emph{0.00} & \emph{0.00} \\
San Jos\'e & Hist.\ GB & 0.75 & [0.72, 0.78] & $-0.07$ & 0.50 & 0.50 \\
 & \emph{Majority baseline} & \emph{0.49} & \emph{--} & \emph{--} & \emph{0.00} & \emph{0.00} \\
Santo Domingo & XGBoost & 0.92 & [0.88, 0.96] & $-0.02$ & 0.84 & 0.86 \\
 & \emph{Majority baseline} & \emph{0.40} & \emph{--} & \emph{--} & \emph{0.00} & \emph{0.00} \\
\bottomrule
\end{tabular}
\end{table}

\begin{figure}[H]
\centering
\includegraphics[width=0.71\textwidth]{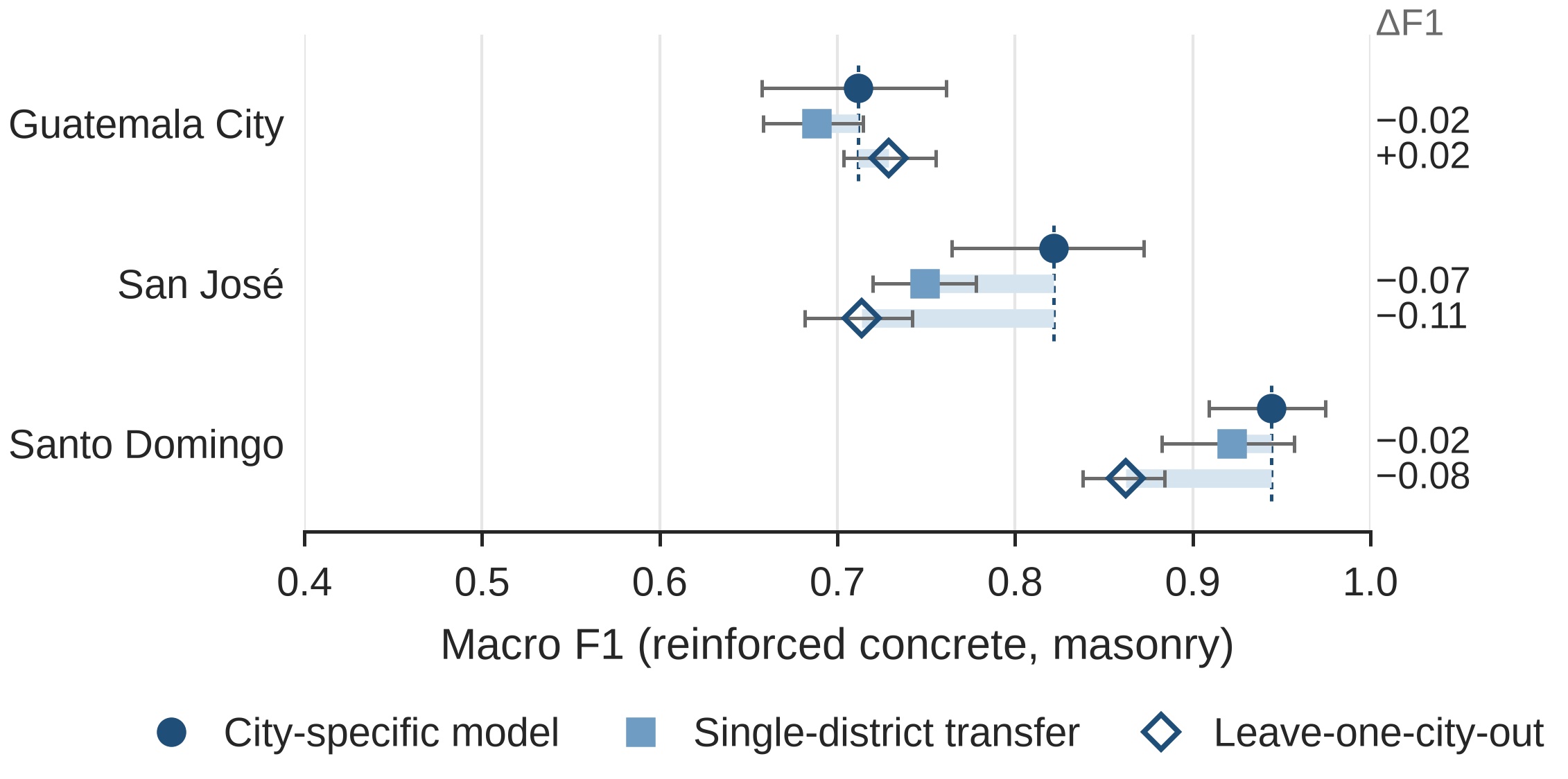}
\caption{Macro F1 on reinforced concrete and masonry of the best city-specific and transferred models (Tables~\ref{tab:basecities} and~\ref{tab:transfer}), with 95\% block-bootstrap CIs. Dashed line: city-specific model.}
\label{fig:transfer-loss}
\end{figure}

Both transfer scenarios outperform their majority-class baselines in all three cities (Table~\ref{tab:transfer}), and five of the six fall below the best within-city model on the classes they share (Figure~\ref{fig:transfer-loss}). Cross-city transfer to Guatemala City scores slightly above it ($\Delta\text{F1} = +0.02$). The difference is within the city-specific model's confidence interval for both Guatemala City scenarios and for district transfer in Santo Domingo, and the loss is largest for cross-city transfer to San Jos\'e ($\Delta\text{F1} = -0.11$). Cross-city $\kappa$ ranges from 0.26 in San Jos\'e to 0.72 in Santo Domingo. This is not a ranking of transferability, because classes absent from the training data are also dropped from the target (adobe in Guatemala City, S/W in San Jos\'e). A transferred model cannot predict such locally specific building types at all, so a city-specific model remains preferable whenever local labels are available.\\
Within-city district transfer exhibits a different pattern. In Guatemala City,
the district-transfer performance is lower than cross-city performance 
($\kappa=0.38$ versus $0.46$), while the reverse is observed in San Jos\'e (0.50 versus 0.26). Santo Domingo achieves comparatively high performance in both scenarios, with $\kappa=0.84$ for district transfer and $0.72$ for cross-city transfer, although the district-transfer estimate is based on a small test set of 211 buildings. These contrasting patterns indicate that cross-city performance alone does not reliably characterize transferability between districts, supporting the need for a separate assessment of within-city transfer.

Without target labels, disagreement among the five models could indicate where a transferred model is unreliable. In five of the six transfer settings, the buildings on which the models disagree (6--29\% of each target set) contain 33--77\% of the errors of their majority vote, with error rates of 30--51\% versus 2--7\% where all five models agree (area under the receiver operating characteristic curve, AUC, of the vote entropy as an error detector: 0.65--0.82). The exception is the transfer from Zona~1 to Zona~10, where the models disagree on 76\% of buildings, so disagreement flags most of the target set although it still separates errors from correct predictions (AUC $= 0.67$). Errors shared by all five models remain undetectable by this criterion.

\subsection{Assessment of a single global model for the three cities}
\label{sec:exp-pooled}

We also assess whether a single model trained on the combined data of the three cities can outperform separate city-specific models. This approach would allow a
single model to be updated as additional data become available from new locations. Unlike the 
leave-one-city-out setup in Table~\ref{tab:transfer}, no city is excluded from training. Instead, the labeled buildings of the three cities are combined and partitioned into training, validation, and test sets using a single spatially grouped split, with each city contributing buildings to all three partitions. Test performance is reported separately for each city
in Table~\ref{tab:pooled} to assess the performance of the three-city model in the three urban contexts.

\begin{table}[H]
\centering
\small
\caption{Models trained on the combined data of the three cities. GT, SJ, SD: macro F1 on each city's test buildings, over the classes present. Bold: best per column. 95\% CI (All): percentile bootstrap over 2000 resamples of spatial blocks.}
\label{tab:pooled}
\footnotesize
\begin{tabular}{lrrrrrr}
\toprule
 & \multicolumn{5}{c}{Macro F1} & \\
\cmidrule(lr){2-6}
Model & All & 95\% CI & GT & SJ & SD & $\kappa$ \\
\midrule
CatBoost & 0.57 & [0.50, 0.63] & 0.42 & \textbf{0.69} & 0.90 & 0.56 \\
Hist.\ GB & 0.56 & [0.49, 0.63] & 0.44 & 0.62 & 0.91 & 0.50 \\
Logistic Regression & 0.54 & [0.47, 0.60] & 0.41 & 0.64 & 0.89 & 0.51 \\
Random Forest & 0.58 & [0.49, 0.65] & 0.44 & 0.68 & 0.86 & 0.54 \\
XGBoost & \textbf{0.63} & [0.53, 0.69] & \textbf{0.47} & 0.69 & \textbf{0.95} & 0.60 \\
\emph{Majority baseline} & \emph{0.24} & \emph{--} & \emph{--} & \emph{--} & \emph{--} & \emph{0.00} \\
\bottomrule
\end{tabular}
\end{table}

XGBoost achieves the highest overall macro F1 on the combined three-city test set (0.63, Table~\ref{tab:pooled}). However, this aggregate score is influenced by the unequal contributions of the three cities: Guatemala City and
San Jos\'e account for 971 and 1354 of the 2550 test buildings, respectively, compared with 225 from Santo Domingo. XGBoost also scores highest in Guatemala City (0.47) and Santo Domingo (0.95), whereas CatBoost scores highest in San Jos\'e (0.69), only 0.001 above XGBoost.\\
Compared on the classes shared with the city-specific models, the three-city models remain close to the best city-specific model (Table~\ref{tab:basecities}, using the same test buildings in San Jos\'e and Santo Domingo): 0.55 to 0.63 versus 0.58 in Guatemala City, 0.62 to 0.69 versus 0.71 in San Jos\'e, and 0.86 to 0.95 versus 0.94 in Santo Domingo. Combining the cities therefore costs little, and the three-city XGBoost model even scores slightly above the city-specific model in Guatemala City and Santo Domingo. It offers a practical alternative when maintaining a single model is preferred.

\subsection{Feature importance and attribute ablation}
\label{sec:exp-ablation}
\label{sec:exp-interpretability}

\begin{figure}[H]
\centering
\includegraphics[width=0.69\textwidth]{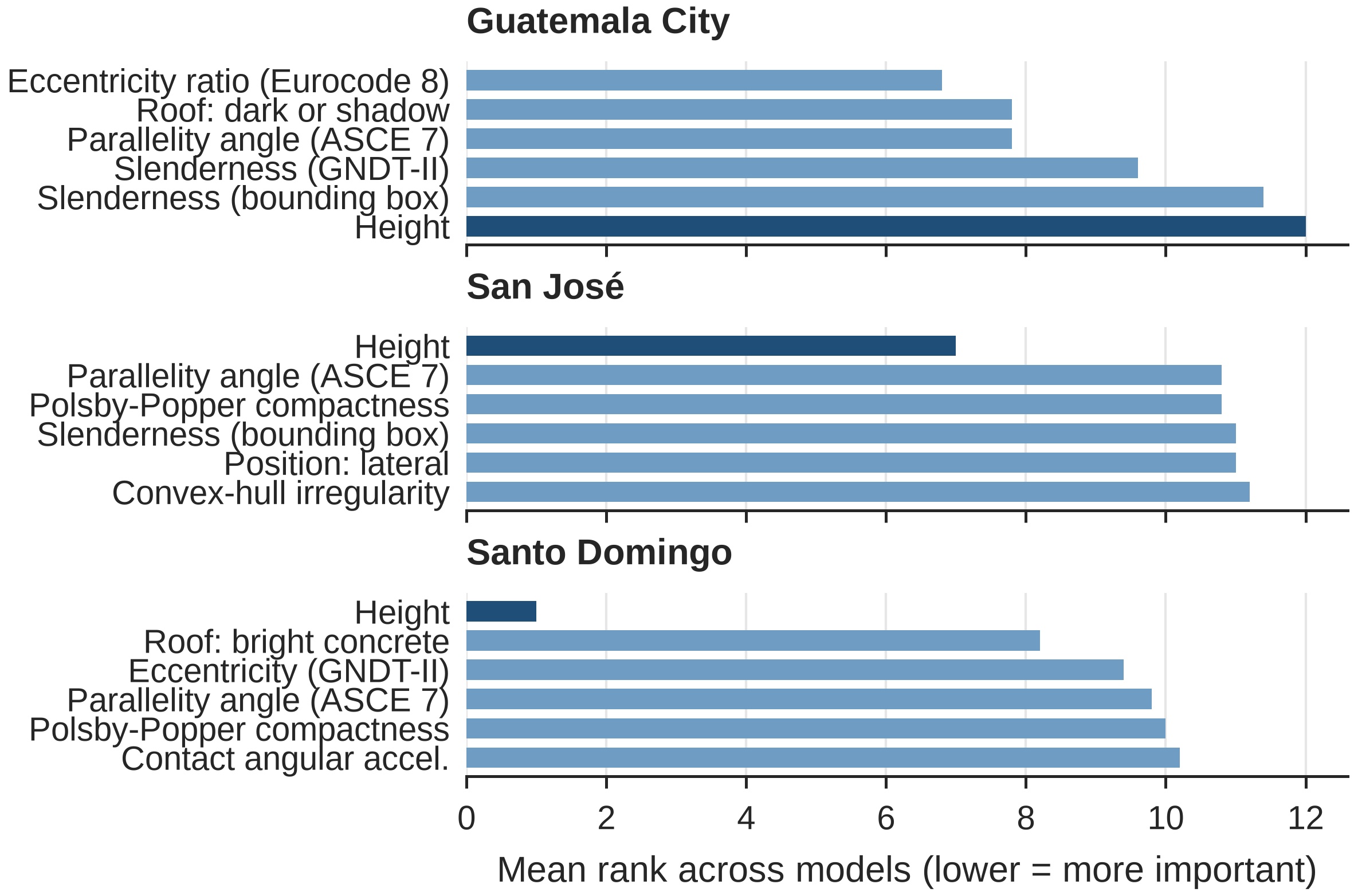}
\caption{Consensus feature ranking per city: six highest-ranked features, lower mean rank is more important. Height in dark blue.}
\label{fig:feature-importance}
\end{figure}

Figure~\ref{fig:feature-importance} shows the consensus feature ranking for each city. Height is first in San Jos\'e and Santo Domingo but sixth in Guatemala City, and the ASCE~7 parallelity angle is the most stable feature, second to fourth in all three cities. The contact confinement ratio ranks 11th to 30th. Other features matter mainly in some cities: the Eurocode~8 eccentricity ratio is first in Guatemala City and eighth in San Jos\'e but last in Santo Domingo, a shape-compactness measure (the Polsby--Popper score) is third in San Jos\'e and fifth in Santo Domingo but 23rd in Guatemala City, and the NTC-23 and ASCE~7 setback ratios range from eighth to 23rd.\\
Among the inexpensive non-geometric attributes, construction era ranks last in Guatemala City, 31st of 32 in San Jos\'e and 29th of 31 in Santo Domingo, code quality ranks 16th to 26th, and roof material ranks second in Guatemala City and Santo Domingo and 14th in San Jos\'e. Retraining each city's best model without these three attributes, on the same test buildings as in Table~\ref{tab:basecities}, changes macro F1 by $+0.011$ in Guatemala City, $-0.014$ in San Jos\'e, and $+0.005$ in Santo Domingo. The 95\% paired block-bootstrap interval includes zero in all three cities. These attributes can therefore be removed with little loss, although the joint ablation does not isolate their individual contributions. Construction era is the clearest candidate for exclusion. Confirming this beyond the three cities requires further surveys.

%% file: sections/conclusion.tex
Classifiers based on building footprint geometry and a few low-cost attributes provide a starting point for structural-system identification in new urban areas, but local reference data remain necessary for validation. Within-city performance varies substantially across the three cities, so performance in one urban context should not be assumed to carry over to another, and it still improves modestly with more local labels in Guatemala City and San Jos\'e, although a few hundred labels already capture most of it where rare classes are well represented. Cross-city and cross-district transfer outperform their majority-class baselines but mostly remain below within-city performance and cannot predict building types absent from their training data. Each transfer setting therefore requires independent validation. A single model trained on the combined data of the three cities performs close to or above the city-specific models in all three cities and offers a practical alternative when maintaining a single model is preferred.\\
Height is the most important predictor in two of the three cities, and accurate height data should therefore be prioritized in future data collection. Jointly excluding roof material, construction era, and a code-quality proxy changes macro F1 by at most 0.02 in any city.\\
These conclusions are subject to four limitations. First, class and location are entangled. Nearly all adobe buildings are in Guatemala City (99\%) and most S/W buildings in San Jos\'e (91\%, Table~\ref{tab:class-imbalance}). A model cannot separate what a class looks like from what buildings in that one place look like when nearly every example comes from there. When that city is held out, the class is absent from training and cannot be predicted at all, and resampling such as SMOTE cannot compensate, because it only interpolates between examples from the same city. The three leave-one-city-out scores therefore cover different class sets and do not rank the cities by transferability.\\
Second, Santo Domingo's strong results reflect an easier task. With 969 labeled buildings split roughly two-to-one between masonry and reinforced concrete, it is the smallest dataset and the only two-class one, whereas the other two cities pose three-class problems with 85\% or more masonry (Table~\ref{tab:class-imbalance}). Its higher scores therefore say more about the favorable shape of its problem than about the pipeline's general ceiling.\\
Third, the remote-sensing attributes carry no uncertainty. Height, construction era, and roof material enter the models as point estimates. GlobalBuildingAtlas reports a height RMSE of 5.5\,m globally (1.5 to 8.9\,m across continents) \citep{zhu2025globalbuildingatlas}, a large fraction of a one- or two-story building's height, and neither the era nor the roof-material product provides a per-building confidence. The models therefore treat every attribute as equally reliable for every building.\\
Finally, the ground truth may contain errors. Even though the surveys were carried out by engineering and architecture students and a significant portion of the labeled buildings has been reviewed, human error is hard to avoid completely, and the structural system is estimated only from the visual appearance of the facade at street level.\\
Before a model is applied to a new district or city, its performance should be assessed on a locally labeled sample. Where no local labels exist, buildings on which the classifiers disagree could be surveyed first, since they concentrate most transfer errors in five of the six settings, although this criterion misses errors shared by all models and selects most buildings where the models disagree widely. Overall, these models support preliminary exposure characterization and targeted review, with local validation remaining essential to inventory development.

%% file: sections/future-work.tex
Several extensions could strengthen these results. Further review of the buildings on which all models disagree with the recorded label would help separate actual class overlap from residual label noise, and calibrated per-building uncertainty could improve on the model-disagreement criterion, which misses errors shared by all models. The class-location entanglement is a data problem rather than a modeling one: surveys in cities where adobe, light steel and wood, or other under-represented classes are common, ideally labeled with the full GEM taxonomy, would let models learn what a class looks like independently of where it was built. Finally, street-level imagery could complement in-person surveys, as \citet{aravenapelizari2021automated} demonstrated with deep learning, and vision-language models could be benchmarked against engineers' assessments of the same images.

%% file: main.bbl
\begin{thebibliography}{43}
\expandafter\ifx\csname natexlab\endcsname\relax\def\natexlab#1{#1}\fi
\providecommand{\url}[1]{\texttt{#1}}
\providecommand{\href}[2]{#2}
\providecommand{\path}[1]{#1}
\providecommand{\DOIprefix}{doi:}
\providecommand{\ArXivprefix}{arXiv:}
\providecommand{\URLprefix}{URL: }
\providecommand{\Pubmedprefix}{pmid:}
\providecommand{\doi}[1]{\href{http://dx.doi.org/#1}{\path{#1}}}
\providecommand{\Pubmed}[1]{\href{pmid:#1}{\path{#1}}}
\providecommand{\bibinfo}[2]{#2}
\ifx\xfnm\relax \def\xfnm[#1]{\unskip,\space#1}\fi
%Type = Article
\bibitem[{Silva et~al.(2022)Silva, Brzev, Scawthorn, Yepes, Dabbeek, and
  Crowley}]{silva2022buildingclassification}
\bibinfo{author}{V.~Silva}, \bibinfo{author}{S.~Brzev},
  \bibinfo{author}{C.~Scawthorn}, \bibinfo{author}{C.~Yepes},
  \bibinfo{author}{J.~Dabbeek}, \bibinfo{author}{H.~Crowley},
\newblock \bibinfo{title}{A building classification system for multi-hazard
  risk assessment},
\newblock \bibinfo{journal}{International Journal of Disaster Risk Science}
  \bibinfo{volume}{13} (\bibinfo{year}{2022}) \bibinfo{pages}{161--177}.
  \DOIprefix\doi{10.1007/s13753-022-00400-x}.
%Type = Article
\bibitem[{Wieland et~al.(2012)Wieland, Pittore, Parolai, and
  Zschau}]{wieland2012exposure}
\bibinfo{author}{M.~Wieland}, \bibinfo{author}{M.~Pittore},
  \bibinfo{author}{S.~Parolai}, \bibinfo{author}{J.~Zschau},
\newblock \bibinfo{title}{Exposure estimation from multi-resolution optical
  satellite imagery for seismic risk assessment},
\newblock \bibinfo{journal}{ISPRS International Journal of Geo-Information}
  \bibinfo{volume}{1} (\bibinfo{year}{2012}) \bibinfo{pages}{69--88}.
  \DOIprefix\doi{10.3390/ijgi1010069}.
%Type = Article
\bibitem[{Pittore et~al.(2016)Pittore, Wieland, and
  Fleming}]{pittore2016exposure}
\bibinfo{author}{M.~Pittore}, \bibinfo{author}{M.~Wieland},
  \bibinfo{author}{K.~Fleming},
\newblock \bibinfo{title}{Perspectives on global dynamic exposure modelling for
  geo-risk assessment},
\newblock \bibinfo{journal}{Natural Hazards} \bibinfo{volume}{86}
  (\bibinfo{year}{2016}) \bibinfo{pages}{7--30}.
  \DOIprefix\doi{10.1007/s11069-016-2437-3}.
%Type = Article
\bibitem[{Torres et~al.(2023)Torres, Mart{\'i}nez-Cuevas, Molina-Palacios,
  Arranz, and Arredondo}]{torres2023remotesensing}
\bibinfo{author}{Y.~Torres}, \bibinfo{author}{S.~Mart{\'i}nez-Cuevas},
  \bibinfo{author}{S.~Molina-Palacios}, \bibinfo{author}{J.~J. Arranz},
  \bibinfo{author}{{\'A}.~Arredondo},
\newblock \bibinfo{title}{Using remote sensing for exposure and seismic
  vulnerability evaluation: is it reliable?},
\newblock \bibinfo{journal}{GIScience \& Remote Sensing} \bibinfo{volume}{60}
  (\bibinfo{year}{2023}) \bibinfo{pages}{2196162}.
  \DOIprefix\doi{10.1080/15481603.2023.2196162}.
%Type = Article
\bibitem[{Yepes-Estrada et~al.(2023)Yepes-Estrada, Calderon, Costa, Crowley,
  Dabbeek, Hoyos, Martins, Paul, Rao, and Silva}]{yepesestrada2023global}
\bibinfo{author}{C.~Yepes-Estrada}, \bibinfo{author}{A.~Calderon},
  \bibinfo{author}{C.~Costa}, \bibinfo{author}{H.~Crowley},
  \bibinfo{author}{J.~Dabbeek}, \bibinfo{author}{M.~C. Hoyos},
  \bibinfo{author}{L.~Martins}, \bibinfo{author}{N.~Paul},
  \bibinfo{author}{A.~Rao}, \bibinfo{author}{V.~Silva},
\newblock \bibinfo{title}{Global building exposure model for earthquake risk
  assessment},
\newblock \bibinfo{journal}{Earthquake Spectra} \bibinfo{volume}{39}
  (\bibinfo{year}{2023}) \bibinfo{pages}{2212--2235}.
  \DOIprefix\doi{10.1177/87552930231194048}.
%Type = Article
\bibitem[{Dabbeek et~al.(2021)Dabbeek, Crowley, Silva, Weatherill, Paul, and
  Nievas}]{dabbeek2021exposure}
\bibinfo{author}{J.~Dabbeek}, \bibinfo{author}{H.~Crowley},
  \bibinfo{author}{V.~Silva}, \bibinfo{author}{G.~Weatherill},
  \bibinfo{author}{N.~Paul}, \bibinfo{author}{C.~I. Nievas},
\newblock \bibinfo{title}{Impact of exposure spatial resolution on seismic loss
  estimates in regional portfolios},
\newblock \bibinfo{journal}{Bulletin of Earthquake Engineering}
  \bibinfo{volume}{19} (\bibinfo{year}{2021}) \bibinfo{pages}{5819--5841}.
  \DOIprefix\doi{10.1007/s10518-021-01194-x}.
%Type = Article
\bibitem[{Rodr{\'i}guez-Saiz et~al.(2025)Rodr{\'i}guez-Saiz,
  Gonz{\'a}lez-Rodrigo, Rejas-Ayuga, Hidalgo-Leiva, and
  Marchamalo-Sacrist{\'a}n}]{rodriguezsaiz2025geospatial}
\bibinfo{author}{J.~Rodr{\'i}guez-Saiz},
  \bibinfo{author}{B.~Gonz{\'a}lez-Rodrigo}, \bibinfo{author}{J.~G.
  Rejas-Ayuga}, \bibinfo{author}{D.~A. Hidalgo-Leiva},
  \bibinfo{author}{M.~Marchamalo-Sacrist{\'a}n},
\newblock \bibinfo{title}{Assessing building seismic exposure using geospatial
  technologies in data-scarce environments: Case study of san jos{\'e}, costa
  rica},
\newblock \bibinfo{journal}{Applied Sciences} \bibinfo{volume}{15}
  (\bibinfo{year}{2025}) \bibinfo{pages}{6318}.
  \DOIprefix\doi{10.3390/app15116318}.
%Type = Article
\bibitem[{Ure{\~n}a-Pliego et~al.(2026)Ure{\~n}a-Pliego, Rodr{\'i}guez-Saiz,
  N{\'u}{\~n}ez-{\'A}lvarez, Marchamalo-Sacrist{\'a}n, and
  Gonz{\'a}lez-Rodrigo}]{urenapliego2026footprint}
\bibinfo{author}{M.~Ure{\~n}a-Pliego}, \bibinfo{author}{J.~Rodr{\'i}guez-Saiz},
  \bibinfo{author}{G.~N{\'u}{\~n}ez-{\'A}lvarez},
  \bibinfo{author}{M.~Marchamalo-Sacrist{\'a}n},
  \bibinfo{author}{B.~Gonz{\'a}lez-Rodrigo},
\newblock \bibinfo{title}{A methodology for the automated estimation of
  footprint-derived seismic behaviour modifiers in building exposure
  assessment},
\newblock \bibinfo{journal}{Advanced Modeling and Simulation in Engineering
  Sciences} \bibinfo{volume}{13} (\bibinfo{year}{2026}) \bibinfo{pages}{3}.
  \DOIprefix\doi{10.1186/s40323-026-00323-y}.
%Type = Techreport
\bibitem[{Brzev et~al.(2013)Brzev, Scawthorn, Charleson, Allen, Greene,
  Jaiswal, and Silva}]{brzev2013gemtaxonomy}
\bibinfo{author}{S.~Brzev}, \bibinfo{author}{C.~Scawthorn},
  \bibinfo{author}{A.~W. Charleson}, \bibinfo{author}{L.~Allen},
  \bibinfo{author}{M.~Greene}, \bibinfo{author}{K.~Jaiswal},
  \bibinfo{author}{V.~Silva}, \bibinfo{title}{{GEM} Building Taxonomy Version
  2.0}, \bibinfo{type}{Technical Report} \bibinfo{number}{GEM Technical Report
  2013-02 V1.0.0}, GEM Foundation, \bibinfo{address}{Pavia, Italy},
  \bibinfo{year}{2013}. \DOIprefix\doi{10.13117/GEM.EXP-MOD.TR2013.02}.
%Type = Article
\bibitem[{Jim{\'e}nez-Mart{\'i}nez et~al.(2024)Jim{\'e}nez-Mart{\'i}nez,
  Navas-S{\'a}nchez, Gonz{\'a}lez-Rodrigo, and
  Hern{\'a}ndez-Rubio}]{jimenezmartinez2024fragility}
\bibinfo{author}{M.~Jim{\'e}nez-Mart{\'i}nez},
  \bibinfo{author}{L.~Navas-S{\'a}nchez},
  \bibinfo{author}{B.~Gonz{\'a}lez-Rodrigo},
  \bibinfo{author}{O.~Hern{\'a}ndez-Rubio},
\newblock \bibinfo{title}{A methodology to assess and select seismic fragility
  curves: Calibration from expert survey and fuzzy analysis},
\newblock \bibinfo{journal}{International Journal of Disaster Risk Reduction}
  \bibinfo{volume}{114} (\bibinfo{year}{2024}) \bibinfo{pages}{104930}.
  \DOIprefix\doi{10.1016/j.ijdrr.2024.104930}.
%Type = Article
\bibitem[{Cardellicchio et~al.(2022)Cardellicchio, Ruggieri, Leggieri, and
  Uva}]{cardellicchio2022vulma}
\bibinfo{author}{A.~Cardellicchio}, \bibinfo{author}{S.~Ruggieri},
  \bibinfo{author}{V.~Leggieri}, \bibinfo{author}{G.~Uva},
\newblock \bibinfo{title}{View {VULMA}: Data set for training a
  machine-learning tool for a fast vulnerability analysis of existing
  buildings},
\newblock \bibinfo{journal}{Data} \bibinfo{volume}{7} (\bibinfo{year}{2022})
  \bibinfo{pages}{4}. \DOIprefix\doi{10.3390/data7010004}.
%Type = Article
\bibitem[{Pittore and Wieland(2013)}]{pittore2013toward}
\bibinfo{author}{M.~Pittore}, \bibinfo{author}{M.~Wieland},
\newblock \bibinfo{title}{Toward a rapid probabilistic seismic vulnerability
  assessment using satellite and ground-based remote sensing},
\newblock \bibinfo{journal}{Natural Hazards} \bibinfo{volume}{68}
  (\bibinfo{year}{2013}) \bibinfo{pages}{115--145}.
  \DOIprefix\doi{10.1007/s11069-012-0475-z}.
%Type = Article
\bibitem[{Gei{\ss} et~al.(2015)Gei{\ss}, Aravena~Pelizari, Marconcini, Sengara,
  Edwards, Lakes, and Taubenb{\"o}ck}]{geiss2015estimation}
\bibinfo{author}{C.~Gei{\ss}}, \bibinfo{author}{P.~Aravena~Pelizari},
  \bibinfo{author}{M.~Marconcini}, \bibinfo{author}{W.~Sengara},
  \bibinfo{author}{M.~Edwards}, \bibinfo{author}{T.~Lakes},
  \bibinfo{author}{H.~Taubenb{\"o}ck},
\newblock \bibinfo{title}{Estimation of seismic building structural types using
  multi-sensor remote sensing and machine learning techniques},
\newblock \bibinfo{journal}{ISPRS Journal of Photogrammetry and Remote Sensing}
  \bibinfo{volume}{104} (\bibinfo{year}{2015}) \bibinfo{pages}{175--188}.
  \DOIprefix\doi{10.1016/j.isprsjprs.2014.07.016}.
%Type = Article
\bibitem[{Riedel et~al.(2015)Riedel, Gu{\'e}guen, Dalla~Mura, Pathier, Leduc,
  and Chanussot}]{riedel2015seismic}
\bibinfo{author}{I.~Riedel}, \bibinfo{author}{P.~Gu{\'e}guen},
  \bibinfo{author}{M.~Dalla~Mura}, \bibinfo{author}{E.~Pathier},
  \bibinfo{author}{T.~Leduc}, \bibinfo{author}{J.~Chanussot},
\newblock \bibinfo{title}{Seismic vulnerability assessment of urban
  environments in moderate-to-low seismic hazard regions using association rule
  learning and support vector machine methods},
\newblock \bibinfo{journal}{Natural Hazards} \bibinfo{volume}{76}
  (\bibinfo{year}{2015}) \bibinfo{pages}{1111--1141}.
  \DOIprefix\doi{10.1007/s11069-014-1538-0}.
%Type = Article
\bibitem[{Aravena~Pelizari et~al.(2021)Aravena~Pelizari, Gei{\ss}, Aguirre,
  Santa~Mar{\'i}a, Merino~Pe{\~n}a, and
  Taubenb{\"o}ck}]{aravenapelizari2021automated}
\bibinfo{author}{P.~Aravena~Pelizari}, \bibinfo{author}{C.~Gei{\ss}},
  \bibinfo{author}{P.~Aguirre}, \bibinfo{author}{H.~Santa~Mar{\'i}a},
  \bibinfo{author}{Y.~Merino~Pe{\~n}a}, \bibinfo{author}{H.~Taubenb{\"o}ck},
\newblock \bibinfo{title}{Automated building characterization for seismic risk
  assessment using street-level imagery and deep learning},
\newblock \bibinfo{journal}{ISPRS Journal of Photogrammetry and Remote Sensing}
  \bibinfo{volume}{180} (\bibinfo{year}{2021}) \bibinfo{pages}{370--386}.
  \DOIprefix\doi{10.1016/j.isprsjprs.2021.07.004}.
%Type = Article
\bibitem[{Liuzzi et~al.(2019)Liuzzi, Aravena~Pelizari, Gei{\ss}, Masi,
  Tramutoli, and Taubenb{\"o}ck}]{liuzzi2019transferable}
\bibinfo{author}{M.~Liuzzi}, \bibinfo{author}{P.~Aravena~Pelizari},
  \bibinfo{author}{C.~Gei{\ss}}, \bibinfo{author}{A.~Masi},
  \bibinfo{author}{V.~Tramutoli}, \bibinfo{author}{H.~Taubenb{\"o}ck},
\newblock \bibinfo{title}{A transferable remote sensing approach to classify
  building structural types for seismic risk analyses: The case of val d'agri
  area (italy)},
\newblock \bibinfo{journal}{Bulletin of Earthquake Engineering}
  \bibinfo{volume}{17} (\bibinfo{year}{2019}) \bibinfo{pages}{4825--4853}.
  \DOIprefix\doi{10.1007/s10518-019-00648-7}.
%Type = Article
\bibitem[{Allouche et~al.(2006)Allouche, Tsoar, and
  Kadmon}]{allouche2006assessing}
\bibinfo{author}{O.~Allouche}, \bibinfo{author}{A.~Tsoar},
  \bibinfo{author}{R.~Kadmon},
\newblock \bibinfo{title}{Assessing the accuracy of species distribution
  models: Prevalence, kappa and the true skill statistic (tss)},
\newblock \bibinfo{journal}{Journal of Applied Ecology} \bibinfo{volume}{43}
  (\bibinfo{year}{2006}) \bibinfo{pages}{1223--1232}.
  \DOIprefix\doi{10.1111/j.1365-2664.2006.01214.x}.
%Type = Article
\bibitem[{Chawla et~al.(2002)Chawla, Bowyer, Hall, and
  Kegelmeyer}]{chawla2002smote}
\bibinfo{author}{N.~V. Chawla}, \bibinfo{author}{K.~W. Bowyer},
  \bibinfo{author}{L.~O. Hall}, \bibinfo{author}{W.~P. Kegelmeyer},
\newblock \bibinfo{title}{{SMOTE}: Synthetic minority over-sampling technique},
\newblock \bibinfo{journal}{Journal of Artificial Intelligence Research}
  \bibinfo{volume}{16} (\bibinfo{year}{2002}) \bibinfo{pages}{321--357}.
  \DOIprefix\doi{10.1613/jair.953}.
%Type = Article
\bibitem[{Steiniger et~al.(2008)Steiniger, Lange, Burghardt, and
  Weibel}]{steiniger2008classification}
\bibinfo{author}{S.~Steiniger}, \bibinfo{author}{T.~Lange},
  \bibinfo{author}{D.~Burghardt}, \bibinfo{author}{R.~Weibel},
\newblock \bibinfo{title}{An approach for the classification of urban building
  structures based on discriminant analysis techniques},
\newblock \bibinfo{journal}{Transactions in GIS} \bibinfo{volume}{12}
  (\bibinfo{year}{2008}) \bibinfo{pages}{31--59}.
  \DOIprefix\doi{10.1111/j.1467-9671.2008.01085.x}.
%Type = Article
\bibitem[{Hecht et~al.(2015)Hecht, Meinel, and
  Buchroithner}]{hecht2015buildingtypes}
\bibinfo{author}{R.~Hecht}, \bibinfo{author}{G.~Meinel},
  \bibinfo{author}{M.~Buchroithner},
\newblock \bibinfo{title}{Automatic identification of building types based on
  topographic databases -- a comparison of different data sources},
\newblock \bibinfo{journal}{International Journal of Cartography}
  \bibinfo{volume}{1} (\bibinfo{year}{2015}) \bibinfo{pages}{18--31}.
  \DOIprefix\doi{10.1080/23729333.2015.1055644}.
%Type = Article
\bibitem[{Henn et~al.(2012)Henn, R{\"o}mer, Gr{\"o}ger, and
  Pl{\"u}mer}]{henn2012automatic}
\bibinfo{author}{A.~Henn}, \bibinfo{author}{C.~R{\"o}mer},
  \bibinfo{author}{G.~Gr{\"o}ger}, \bibinfo{author}{L.~Pl{\"u}mer},
\newblock \bibinfo{title}{Automatic classification of building types in {3D}
  city models: Using {SVMs} for semantic enrichment of low resolution building
  data},
\newblock \bibinfo{journal}{GeoInformatica} \bibinfo{volume}{16}
  (\bibinfo{year}{2012}) \bibinfo{pages}{281--306}.
  \DOIprefix\doi{10.1007/s10707-011-0131-x}.
%Type = Incollection
\bibitem[{Fan et~al.(2014)Fan, Zipf, and Fu}]{fan2014estimation}
\bibinfo{author}{H.~Fan}, \bibinfo{author}{A.~Zipf}, \bibinfo{author}{Q.~Fu},
\newblock \bibinfo{title}{Estimation of building types on {OpenStreetMap} based
  on urban morphology analysis},
\newblock in: \bibinfo{editor}{J.~Huerta}, \bibinfo{editor}{S.~Schade},
  \bibinfo{editor}{C.~Granell} (Eds.), \bibinfo{booktitle}{Connecting a Digital
  Europe Through Location and Place}, Lecture Notes in Geoinformation and
  Cartography, \bibinfo{publisher}{Springer}, \bibinfo{year}{2014}, pp.
  \bibinfo{pages}{19--35}. \DOIprefix\doi{10.1007/978-3-319-03611-3_2}.
%Type = Article
\bibitem[{Atwal et~al.(2022)Atwal, Anderson, Pfoser, and
  Z{\"u}fle}]{atwal2022predicting}
\bibinfo{author}{K.~S. Atwal}, \bibinfo{author}{T.~Anderson},
  \bibinfo{author}{D.~Pfoser}, \bibinfo{author}{A.~Z{\"u}fle},
\newblock \bibinfo{title}{Predicting building types using {OpenStreetMap}},
\newblock \bibinfo{journal}{Scientific Reports} \bibinfo{volume}{12}
  (\bibinfo{year}{2022}) \bibinfo{pages}{19976}.
  \DOIprefix\doi{10.1038/s41598-022-24263-w}.
%Type = Article
\bibitem[{Salazar et~al.(2022)Salazar, Garland, Ochoa, and
  Pyrcz}]{salazar2022fairsplit}
\bibinfo{author}{J.~Salazar}, \bibinfo{author}{L.~Garland},
  \bibinfo{author}{J.~Ochoa}, \bibinfo{author}{M.~J. Pyrcz},
\newblock \bibinfo{title}{Fair train-test split in machine learning: Mitigating
  spatial autocorrelation for improved prediction accuracy},
\newblock \bibinfo{journal}{Journal of Petroleum Science and Engineering}
  \bibinfo{volume}{209} (\bibinfo{year}{2022}) \bibinfo{pages}{109885}.
  \DOIprefix\doi{10.1016/j.petrol.2021.109885}.
%Type = Article
\bibitem[{Wang et~al.(2023)Wang, Khodadadzadeh, and
  Zurita-Milla}]{wang2023spatialplus}
\bibinfo{author}{Y.~Wang}, \bibinfo{author}{M.~Khodadadzadeh},
  \bibinfo{author}{R.~Zurita-Milla},
\newblock \bibinfo{title}{Spatial+: A new cross-validation method to evaluate
  geospatial machine learning models},
\newblock \bibinfo{journal}{International Journal of Applied Earth Observation
  and Geoinformation} \bibinfo{volume}{121} (\bibinfo{year}{2023})
  \bibinfo{pages}{103364}. \DOIprefix\doi{10.1016/j.jag.2023.103364}.
%Type = Misc
\bibitem[{Gonz\'{a}lez-Rodrigo et~al.(2024)Gonz\'{a}lez-Rodrigo, Hidalgo-Leiva,
  Flores, Germoso, Jim\'{e}nez-Mart\'{\i}nez, Navas-S\'{a}nchez, Orta,
  Tarque-Ruiz, Hern\'{a}ndez-Rubio, Marchamalo-Sacrist\'{a}n, and
  {others}}]{gonzalez2024sustainable}
\bibinfo{author}{B.~Gonz\'{a}lez-Rodrigo}, \bibinfo{author}{D.~Hidalgo-Leiva},
  \bibinfo{author}{O.~Flores}, \bibinfo{author}{C.~Germoso},
  \bibinfo{author}{M.~Jim\'{e}nez-Mart\'{\i}nez},
  \bibinfo{author}{L.~Navas-S\'{a}nchez}, \bibinfo{author}{B.~Orta},
  \bibinfo{author}{N.~Tarque-Ruiz}, \bibinfo{author}{O.~Hern\'{a}ndez-Rubio},
  \bibinfo{author}{M.~Marchamalo-Sacrist\'{a}n}, \bibinfo{author}{{others}},
  \bibinfo{title}{Sustainable and resilient construction in {Central} {America}
  and the {Caribbean} in the face of seismic hazard: {Regional} cooperation
  based on the experience of {Costa} {Rica} (9063-03)}, \bibinfo{year}{2024}.
  \URLprefix \url{https://sismica.lis.ucr.ac.cr/}.
%Type = Article
\bibitem[{D{\'a}vila et~al.(2023)D{\'a}vila, Cabrero, and
  Flores}]{davila2023guatemala}
\bibinfo{author}{L.~D. D{\'a}vila}, \bibinfo{author}{J.~M. Cabrero},
  \bibinfo{author}{O.~Flores},
\newblock \bibinfo{title}{New seismic exposure modeling approach for {Guatemala
  City}},
\newblock \bibinfo{journal}{Bulletin of Earthquake Engineering}
  \bibinfo{volume}{21} (\bibinfo{year}{2023}) \bibinfo{pages}{739--766}.
  \DOIprefix\doi{10.1007/s10518-022-01570-1}.
%Type = Misc
\bibitem[{Esquivel-Salas et~al.(2021)Esquivel-Salas, Pittore, Schmidt-D{\'i}az,
  and Haas}]{esquivelsalas2021sanjose}
\bibinfo{author}{L.~Esquivel-Salas}, \bibinfo{author}{M.~Pittore},
  \bibinfo{author}{V.~Schmidt-D{\'i}az}, \bibinfo{author}{M.~Haas},
  \bibinfo{title}{Photographs of buildings from {San Jos\'e} canton, {Costa
  Rica}, taken from the street}, \bibinfo{howpublished}{4TU.ResearchData,
  dataset collection}, \bibinfo{year}{2021}.
  \DOIprefix\doi{10.4121/c.5618230.v1}.
%Type = Article
\bibitem[{Zhu et~al.(2025)Zhu, Chen, Zhang, Shi, and
  Wang}]{zhu2025globalbuildingatlas}
\bibinfo{author}{X.~X. Zhu}, \bibinfo{author}{S.~Chen},
  \bibinfo{author}{F.~Zhang}, \bibinfo{author}{Y.~Shi},
  \bibinfo{author}{Y.~Wang},
\newblock \bibinfo{title}{{GlobalBuildingAtlas}: an open global and complete
  dataset of building polygons, heights and {LoD1} {3D} models},
\newblock \bibinfo{journal}{Earth System Science Data} \bibinfo{volume}{17}
  (\bibinfo{year}{2025}) \bibinfo{pages}{6647--6668}.
  \DOIprefix\doi{10.5194/essd-17-6647-2025}.
%Type = Article
\bibitem[{Marconcini et~al.(2021)Marconcini, Metz-Marconcini, Esch, and
  Gorelick}]{marconcini2021wsfevolution}
\bibinfo{author}{M.~Marconcini}, \bibinfo{author}{A.~Metz-Marconcini},
  \bibinfo{author}{T.~Esch}, \bibinfo{author}{N.~Gorelick},
\newblock \bibinfo{title}{Understanding current trends in global urbanisation
  -- the world settlement footprint suite},
\newblock \bibinfo{journal}{GI\_Forum} \bibinfo{volume}{9}
  (\bibinfo{year}{2021}) \bibinfo{pages}{33--38}.
  \DOIprefix\doi{10.1553/giscience2021_01_s33}.
%Type = Article
\bibitem[{Pesaresi et~al.(2024)Pesaresi, Schiavina, Politis, Freire,
  Krasnod{\k{e}}bska, Uhl, Carioli, Corbane, Dijkstra, Florio, Friedrich, Gao,
  Leyk, Lu, Maffenini, Mari-Rivero, Melchiorri, Syrris, Van Den~Hoek, and
  Kemper}]{pesaresi2024ghsl}
\bibinfo{author}{M.~Pesaresi}, \bibinfo{author}{M.~Schiavina},
  \bibinfo{author}{P.~Politis}, \bibinfo{author}{S.~Freire},
  \bibinfo{author}{K.~Krasnod{\k{e}}bska}, \bibinfo{author}{J.~H. Uhl},
  \bibinfo{author}{A.~Carioli}, \bibinfo{author}{C.~Corbane},
  \bibinfo{author}{L.~Dijkstra}, \bibinfo{author}{P.~Florio},
  \bibinfo{author}{H.~K. Friedrich}, \bibinfo{author}{J.~Gao},
  \bibinfo{author}{S.~Leyk}, \bibinfo{author}{L.~Lu},
  \bibinfo{author}{L.~Maffenini}, \bibinfo{author}{I.~Mari-Rivero},
  \bibinfo{author}{M.~Melchiorri}, \bibinfo{author}{V.~Syrris},
  \bibinfo{author}{J.~Van Den~Hoek}, \bibinfo{author}{T.~Kemper},
\newblock \bibinfo{title}{Advances on the global human settlement layer by
  joint assessment of {Earth} observation and population survey data},
\newblock \bibinfo{journal}{International Journal of Digital Earth}
  \bibinfo{volume}{17} (\bibinfo{year}{2024}) \bibinfo{pages}{2390454}.
  \DOIprefix\doi{10.1080/17538947.2024.2390454}.
%Type = Article
\bibitem[{Roberts et~al.(2017)Roberts, Bahn, Ciuti, Boyce, Elith,
  Guillera-Arroita, Hauenstein, Lahoz-Monfort, Schr{\"o}der, Thuiller, Warton,
  Wintle, Hartig, and Dormann}]{roberts2017crossvalidation}
\bibinfo{author}{D.~R. Roberts}, \bibinfo{author}{V.~Bahn},
  \bibinfo{author}{S.~Ciuti}, \bibinfo{author}{M.~S. Boyce},
  \bibinfo{author}{J.~Elith}, \bibinfo{author}{G.~Guillera-Arroita},
  \bibinfo{author}{S.~Hauenstein}, \bibinfo{author}{J.~J. Lahoz-Monfort},
  \bibinfo{author}{B.~Schr{\"o}der}, \bibinfo{author}{W.~Thuiller},
  \bibinfo{author}{D.~I. Warton}, \bibinfo{author}{B.~A. Wintle},
  \bibinfo{author}{F.~Hartig}, \bibinfo{author}{C.~F. Dormann},
\newblock \bibinfo{title}{Cross-validation strategies for data with temporal,
  spatial, hierarchical, or phylogenetic structure},
\newblock \bibinfo{journal}{Ecography} \bibinfo{volume}{40}
  (\bibinfo{year}{2017}) \bibinfo{pages}{913--929}.
  \DOIprefix\doi{10.1111/ecog.02881}.
%Type = Article
\bibitem[{He and Garcia(2009)}]{he2009imbalanced}
\bibinfo{author}{H.~He}, \bibinfo{author}{E.~A. Garcia},
\newblock \bibinfo{title}{Learning from imbalanced data},
\newblock \bibinfo{journal}{IEEE Transactions on Knowledge and Data
  Engineering} \bibinfo{volume}{21} (\bibinfo{year}{2009})
  \bibinfo{pages}{1263--1284}. \DOIprefix\doi{10.1109/TKDE.2008.239}.
%Type = Article
\bibitem[{Tomek(1976)}]{tomek1976cnn}
\bibinfo{author}{I.~Tomek},
\newblock \bibinfo{title}{Two modifications of {CNN}},
\newblock \bibinfo{journal}{IEEE Transactions on Systems, Man, and Cybernetics}
  \bibinfo{volume}{SMC-6} (\bibinfo{year}{1976}) \bibinfo{pages}{769--772}.
  \DOIprefix\doi{10.1109/TSMC.1976.4309452}.
%Type = Article
\bibitem[{Lemaitre et~al.(2017)Lemaitre, Nogueira, and
  Aridas}]{lemaitre2017imblearn}
\bibinfo{author}{G.~Lemaitre}, \bibinfo{author}{F.~Nogueira},
  \bibinfo{author}{C.~K. Aridas},
\newblock \bibinfo{title}{Imbalanced-learn: A {Python} toolbox to tackle the
  curse of imbalanced datasets in machine learning},
\newblock \bibinfo{journal}{Journal of Machine Learning Research}
  \bibinfo{volume}{18} (\bibinfo{year}{2017}) \bibinfo{pages}{1--5}.
%Type = Book
\bibitem[{Hastie et~al.(2009)Hastie, Tibshirani, and Friedman}]{hastie2009esl}
\bibinfo{author}{T.~Hastie}, \bibinfo{author}{R.~Tibshirani},
  \bibinfo{author}{J.~Friedman}, \bibinfo{title}{The Elements of Statistical
  Learning: Data Mining, Inference, and Prediction}, \bibinfo{edition}{2nd}
  ed., \bibinfo{publisher}{Springer}, \bibinfo{year}{2009}.
%Type = Techreport
\bibitem[{Chen et~al.(2004)Chen, Liaw, and Breiman}]{chen2004balancedrf}
\bibinfo{author}{C.~Chen}, \bibinfo{author}{A.~Liaw},
  \bibinfo{author}{L.~Breiman}, \bibinfo{title}{Using Random Forest to Learn
  Imbalanced Data}, \bibinfo{type}{Technical Report} \bibinfo{number}{666},
  Department of Statistics, University of California, Berkeley,
  \bibinfo{year}{2004}.
%Type = Inproceedings
\bibitem[{Chen and Guestrin(2016)}]{chen2016xgboost}
\bibinfo{author}{T.~Chen}, \bibinfo{author}{C.~Guestrin},
\newblock \bibinfo{title}{{XGBoost}: A scalable tree boosting system},
\newblock in: \bibinfo{booktitle}{Proceedings of the 22nd ACM SIGKDD
  International Conference on Knowledge Discovery and Data Mining},
  \bibinfo{publisher}{ACM}, \bibinfo{address}{San Francisco, CA, USA},
  \bibinfo{year}{2016}, pp. \bibinfo{pages}{785--794}.
  \DOIprefix\doi{10.1145/2939672.2939785}.
%Type = Inproceedings
\bibitem[{Prokhorenkova et~al.(2018)Prokhorenkova, Gusev, Vorobev, Dorogush,
  and Gulin}]{prokhorenkova2018catboost}
\bibinfo{author}{L.~Prokhorenkova}, \bibinfo{author}{G.~Gusev},
  \bibinfo{author}{A.~Vorobev}, \bibinfo{author}{A.~V. Dorogush},
  \bibinfo{author}{A.~Gulin},
\newblock \bibinfo{title}{{CatBoost}: Unbiased boosting with categorical
  features},
\newblock in: \bibinfo{booktitle}{Advances in Neural Information Processing
  Systems 31 (NeurIPS 2018)}, \bibinfo{year}{2018}, pp.
  \bibinfo{pages}{6639--6649}.
%Type = Inproceedings
\bibitem[{Ke et~al.(2017)Ke, Meng, Finley, Wang, Chen, Ma, Ye, and
  Liu}]{ke2017lightgbm}
\bibinfo{author}{G.~Ke}, \bibinfo{author}{Q.~Meng},
  \bibinfo{author}{T.~Finley}, \bibinfo{author}{T.~Wang},
  \bibinfo{author}{W.~Chen}, \bibinfo{author}{W.~Ma}, \bibinfo{author}{Q.~Ye},
  \bibinfo{author}{T.-Y. Liu},
\newblock \bibinfo{title}{{LightGBM}: A highly efficient gradient boosting
  decision tree},
\newblock in: \bibinfo{booktitle}{Advances in Neural Information Processing
  Systems 30 (NeurIPS 2017)}, \bibinfo{year}{2017}, pp.
  \bibinfo{pages}{3146--3154}.
%Type = Article
\bibitem[{Sokolova and Lapalme(2009)}]{sokolova2009systematic}
\bibinfo{author}{M.~Sokolova}, \bibinfo{author}{G.~Lapalme},
\newblock \bibinfo{title}{A systematic analysis of performance measures for
  classification tasks},
\newblock \bibinfo{journal}{Information Processing \& Management}
  \bibinfo{volume}{45} (\bibinfo{year}{2009}) \bibinfo{pages}{427--437}.
  \DOIprefix\doi{10.1016/j.ipm.2009.03.002}.
%Type = Article
\bibitem[{Cohen(1960)}]{cohen1960coefficient}
\bibinfo{author}{J.~Cohen},
\newblock \bibinfo{title}{A coefficient of agreement for nominal scales},
\newblock \bibinfo{journal}{Educational and Psychological Measurement}
  \bibinfo{volume}{20} (\bibinfo{year}{1960}) \bibinfo{pages}{37--46}.
  \DOIprefix\doi{10.1177/001316446002000104}.
%Type = Inproceedings
\bibitem[{Lundberg and Lee(2017)}]{lundberg2017shap}
\bibinfo{author}{S.~M. Lundberg}, \bibinfo{author}{S.-I. Lee},
\newblock \bibinfo{title}{A unified approach to interpreting model
  predictions},
\newblock in: \bibinfo{booktitle}{Advances in Neural Information Processing
  Systems 30 (NIPS 2017)}, \bibinfo{year}{2017}, pp.
  \bibinfo{pages}{4765--4774}.

\end{thebibliography}
